\documentclass[twocolumn, aps,nopreprintnumbers,nolongbibliography,prb,10pt,showpacs,showkeys]{revtex4-1}
\usepackage{bm}
\usepackage{textcomp}
\usepackage{hyperref}
\usepackage{url}
\usepackage{amsmath}
\usepackage{amssymb}
\usepackage{amsxtra}
\usepackage{amscd}
\usepackage{amsthm}
\usepackage{amsfonts}
\usepackage{eucal}
\usepackage{latexsym,amsthm}
\usepackage{hhline}
\usepackage{pstricks}
\usepackage{color}
\usepackage{graphicx}

\newcommand{\nc}{\newcommand}
\nc{\on}{\operatorname}
\nc{\wt}{\widetilde}
\nc{\Wick}{{\mathbb :}}
\nc{\R}{{\mathbb R}}

\newcommand{\beq}{\begin{equation}}
\newcommand{\eeq}{\end{equation}}
\newcommand{\bmul}{\begin{multline}}
\newcommand{\emul}{{\end{multline}}}
\newcommand\beqa{\begin{eqnarray}}
\newcommand\eeqa{\end{eqnarray}}
\newcommand\bea{\begin{array}}
\newcommand\eea{\end{array}}
\newcommand\ba{\begin{array}}
\newcommand\ea{\end{array}}
\newcommand{\nn}{\nonumber}

\newcommand{\neqa}{\nonumber\end{eqnarray}}

\newcommand{\eq}[1]{Eq.(\ref{#1})}

\renewcommand{\d}{\partial}

\newcommand{\<}{{\langle}}
\renewcommand{\>}{{\rangle}}

\nc{\CH}{{\mathcal H}}
\nc{\Db}{{\bar D}}
\nc\comment[1]{}

\nc{\CM}{{\mathcal M}}
\nc{\CN}{{\mathcal N}}

\newcommand{\re}{\relax{\rm I\kern-.18em R}}

\nc{\meV}{{\mathrm{\,meV}}}
\nc{\cG}{{\mathcal G}}

\renewcommand{\bar}{\overline} 

\nc{\al}{{\alpha}}

\def\eps{{\epsilon}}
\def\sign{{\rm \, sign }}

\renewcommand{\)}{\right)}
\renewcommand{\(}{\left(}

\begin{document}
\title{Bound states of charges on top of graphene in magnetic field}
\author{Sergey Slizovskiy}
\email{S.Slizovskiy@lboro.ac.uk}
\affiliation{Department of Physics,  Loughborough University,\\
Loughborough LE11 3TU, UK}
\affiliation{NRC ``Kurchatov Institute'' PNPI, Russia}
\begin{abstract}
We show theoretically that in the external magnetic field 
like charges on top of graphene monolayer may be mutually attracted to form macro-molecules. For this to happen graphene needs to be
in Quantum Hall plateau state with local chemical potential being between the Landau levels. Graphene electron(s) gets localized in the 
middle between 
charges and provides overscreening of Coulomb repulsion between the charges. The size of the resulting macro-molecules is of 
the order of the magnetic length ($\sim 10$ nm for magnetic field 10 T).  
The possible stable 
macro-molecules that unit charges can form on graphene in magnetic field are classified.  The binding survives significant  
temperatures, exceeding mobility barriers for many ionically bond impurities. The influence of possible lattice-scale effects of 
valley-mixing are discussed.

Tuning the doping of graphene or the magnetic field,
the binding of impurities can be turned on and off and the macro-molecule size may be tuned. 
This opens  the  perspective to nanoscopic manipulation of ions on graphene by using magnetic field and gating.  
\end{abstract}

\maketitle
\section{Introduction}
 More than ten years since discovery of graphene \cite{Geim2005}, the first genuinely two-dimensional material,  are marked with huge 
body of research. Its applications range from cancer-drug delivery \cite{Biggs,Biggs1}
to novel electronic devices. Still, many of proposed uses of graphene depend crucially on its interface interactions with other 
compounds and impurities.  
Typically, molecular dynamics and DFT methods \cite{Wehling2009, AlkaliGraphene, LiGraphene, HalogenDFT1,HalogenDFT, BondingAdatoms1}  
are used to model the interaction of graphene with substrate and impurities. 
It was shown that there are two types of impurity bonding to graphene \cite{Wehling2009}: covalently bond impurities (e.g. H, CH$_3$, 
F, OH, O)  and ionically bond impurities (e.g. Na, K, Cs, Cl, Br, I). Covalently bond impurities act similar to defects in graphene 
structure and typically have large mobility barriers,  while ionically bond impurities have low mobility barriers (typically lower
than room temperature) and act as mobile electric charges put on top of graphene \cite{Wehling2009, 
GasMovesFreely1, ReviewGasMovesFreely2}.  

Interaction between impurities has attracted significant experimental and theoretical interest.    It was shown \cite{AbaninClustering08} 
that short-range covalently bond impurities tend to form
bound clusters due to a fermionic Casimir effect with binding energy comparable to room temperature scale at distances below 2 nm. 
Related results were presented in ref. \onlinecite{CheianovKekule} where dilute adatoms were shown to have tendency towards a spatially 
correlated state with a hidden Kekul{\`e} mosaic order.  On the other hand, this effect cannot compete with Coulomb repulsion of like 
charges  if the impurities are charged, as is the case
for ionically bond impurities.       
    
Graphene is known to be exceptionally susceptible to magnetic field with divergent diamagnetic susceptibility at the Dirac point and 
strong non-linear effects in magnetization \cite{ReviewMagnetic, Magnetization}. Magnetic field creates large energy gaps near the Dirac
point that reduce intrinsic screening and help to localize electrons in the external potentials.    

Without magnetic field, graphene will produce non-linear screening of external charges \cite{screeningFogler,screening07}.  When 
magnetic field is applied,
the ordinary screening is significantly suppressed \cite{Pyatkovskiy, Kharitonov}  and electrons become localized. 

 In this work we discuss the interaction between charged impurities near the surface of graphene in the strong magnetic field
and show that in some cases the binding due to electrons in graphene can compete with Coulomb repulsion and stable nano-molecules may 
form.  The motivation for this work comes not only from ionically  bond impurities mentioned above: Epitaxial 
graphene on Si-terminated SiC \cite{EpitaxialReview, EpitaxialKim, EpitaxialSTM, EpitaxialDeadLayer} has an important feature of 
positively-charged donors 
appearing dynamically in the ``dead'' carbon layer just below graphene \cite{TzalenchukPinning, TzalenchukNature, TzalenchukBreakdown}. 
This 
 happens due the dominant effect of quantum capacitance \cite{Kopylov} and  due to strong changes in the density of states near 
the Dirac point  under the influence of magnetic field \cite{TzalenchukPinning, MagneticCharging}.  The process looks as the appearance 
of localized holes in the insulating ``dead'' layer due to electrons transferred to graphene.  
Consequently, the charge transfer might dynamically create extra localized states and extra electrons to keep the system
in a robust $\nu \approx 2$  quantum Hall plateau state.  Since the charge transfer is reversible, the holes below graphene can
be considered on the same footing as dynamical charged impurities.
The process of charge transfer allows us to talk about the well-defined local value of the electron chemical potential.
The chemical potential is chosen to be in the gap between the undisturbed Landau levels.  

Another possible context is to consider the well-known semi-classical picture of Quantum Hall effect \cite{ShklovskiiQHE,Gerhardts} 
(QHE). It assumes smooth (on the scale of magnetic length) external potentials and derives the presence of compressible regions 
where electrons screen everything and the incompressible regions that have exactly integer local filling factors.   For 
non-smooth potentials created by impurities close 
to graphene sheet, the semi-classical approach is not the full truth. The self-consistent semi-classical approach gives a 
smoothly-varying effective potential that could be locally 
treated as a chemical potential for our purpose. Our results imply that there is still some life in the incompressible region due to 
screening and possible over-screening of point charges, and thus the local filling factor of incompressible regions could slightly 
deviate from integer values.  

For definiteness, we consider the positively 
charged ions below,  but exactly the same description applies to the negative ions if one makes a particle-hole transformation and 
reverts 
the sign of chemical potential. 
  
\begin{figure*}[]  

\begin{center}
\includegraphics[scale=0.5]{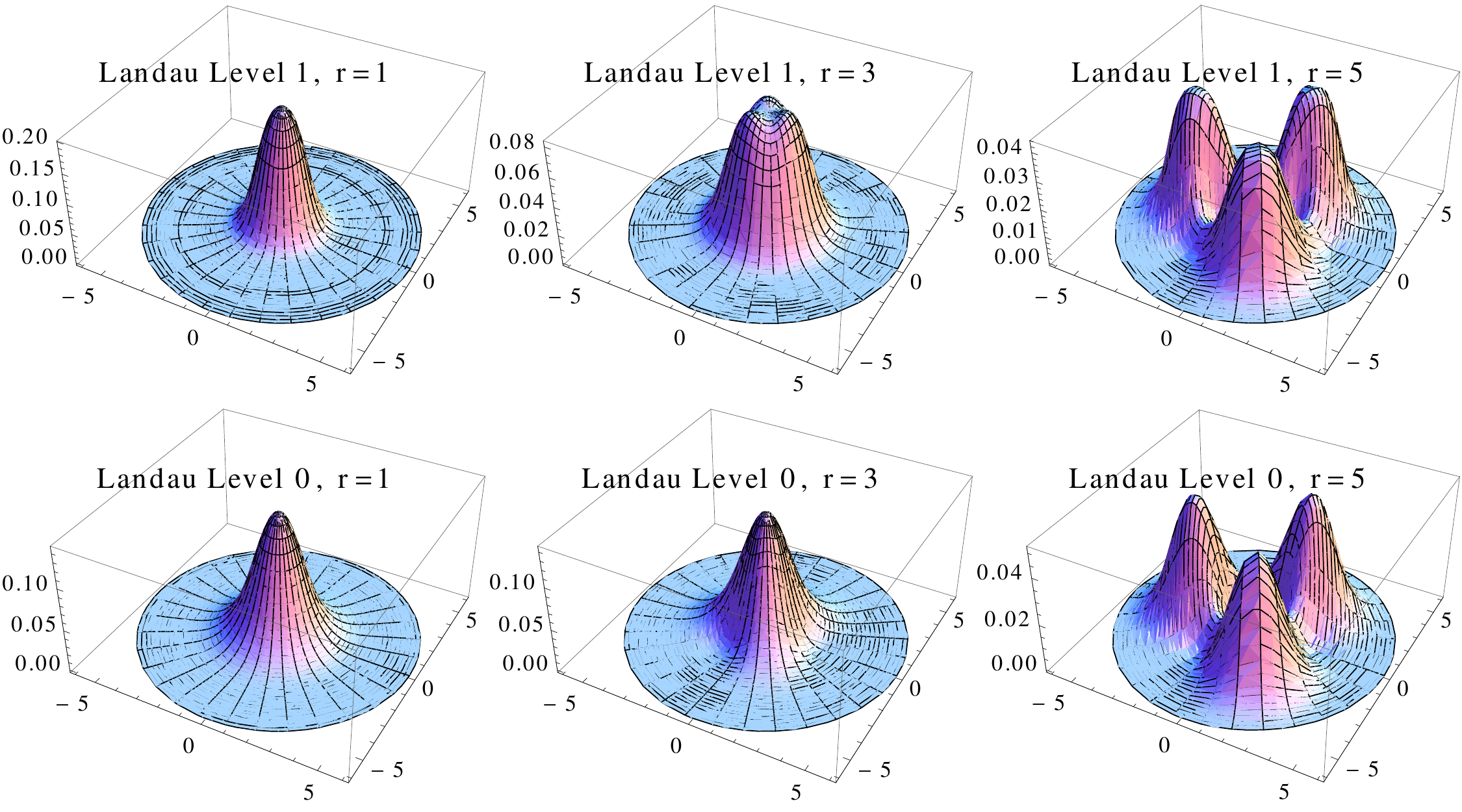}
\end{center}

\vspace{-0.9 cm}
\caption{\label{fig:Wave Function} 
Electron densities for the {\it lowest sublevels} of 0-th and 1-st Landau levels in the field of 3 ions forming equilateral 
triangle with sides $r$. Essential binding occurs for $r\lesssim 3 l_B$ where the electron wave-function forms a single lump in the 
center.
}
\end{figure*}

\begin{figure}[t]  
\begin{center}
\includegraphics[scale=0.5]{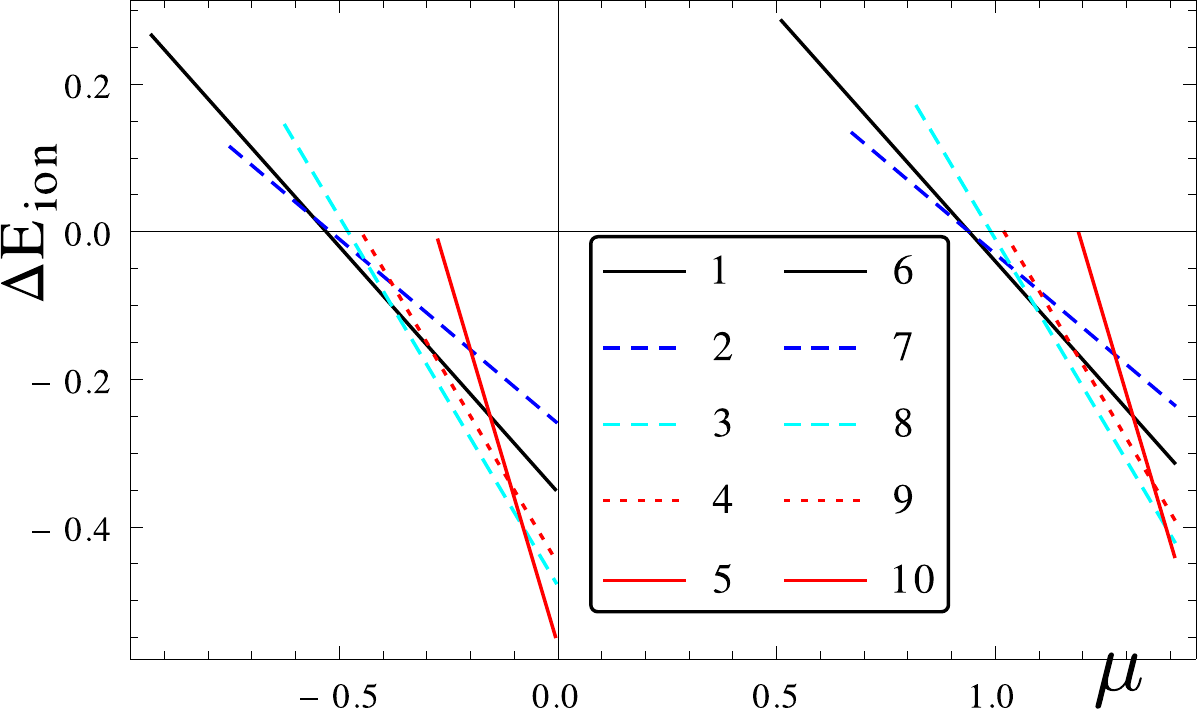}
\end{center}
\vspace{-0.6 cm}
\caption{\label{fig:PhaseDiagram} 
Zero-temperature phase energy plot for dilute mobile charged impurities. Phase with lowest $\Delta E_{\text{ion}}$ is favored. 
Legend corresponds to Table \ref{fig:phase diagram}. The chemical potential should be understood as taken relative to Landau level edges, 
which in this plot are renormalized to free-particle values of $0$ and $\sqrt{2}$ here. }
\end{figure}

We show that when several charged 
impurities are at the distance of the order of magnetic length $l_B=\sqrt{\hbar/(e B)}$, they can form stable molecules bound by the 
electrons in graphene. 
Depending on its local chemical 
potential (taken in the region of Quantum Hall plateu), graphene may bind either positive ions, or negative, or produce no significant 
binding at 
all, the results are summarized 
in Table \ref{fig:phase diagram} and Fig.\ref{fig:PhaseDiagram}. 
The described effect may 
be called either as {\it over-screening of Coulomb repulsion of impurities}  or as {\it long range covalent bonding of charged 
impurities}.  
 A related ``overscreening'' effect exists for
the charging of a quantum dot \cite{ShklovskiiA};  in colloidal systems a similar effect is called
"charge inversion" \cite{ShklovskiiBInversion}.

Graphene in 
magnetic field hosts the electron hybridization cloud that can lead to attraction exceeding the Coulomb 
repulsion of same-charge ions, see Fig.\ref{fig:Wave Function}. Significant binding of charges occurs only when the electron 
cloud is centered in between the charges ($r\lesssim 3$ on Fig.\ref{fig:Wave Function}).   Similarly 
to the signatures of Quantum Hall Effect (QHE) \cite{roomTQHE}, 
we show that the effect may survive the room temperature for magnetic fields of order 10 T.

The paper has the following structure.
In section \ref{sec:1particle} we introduce the formalism and present the results for bound states bound with 1 electron;
in section \ref{sec:multi-electron} the results are generalized to multi-electron bound states;  in section \ref{sec:phase diagram}
the obtained results are combined in a phase diagram; in section \ref{sec:Valley Mixing} the qualitative effects of  potentials that are 
non-smooth on the lattice scale are discussed; conclusions are given in section \ref{sec:conclusions}.


\begin{figure}[t]  
\begin{center}
\includegraphics[scale=0.5]{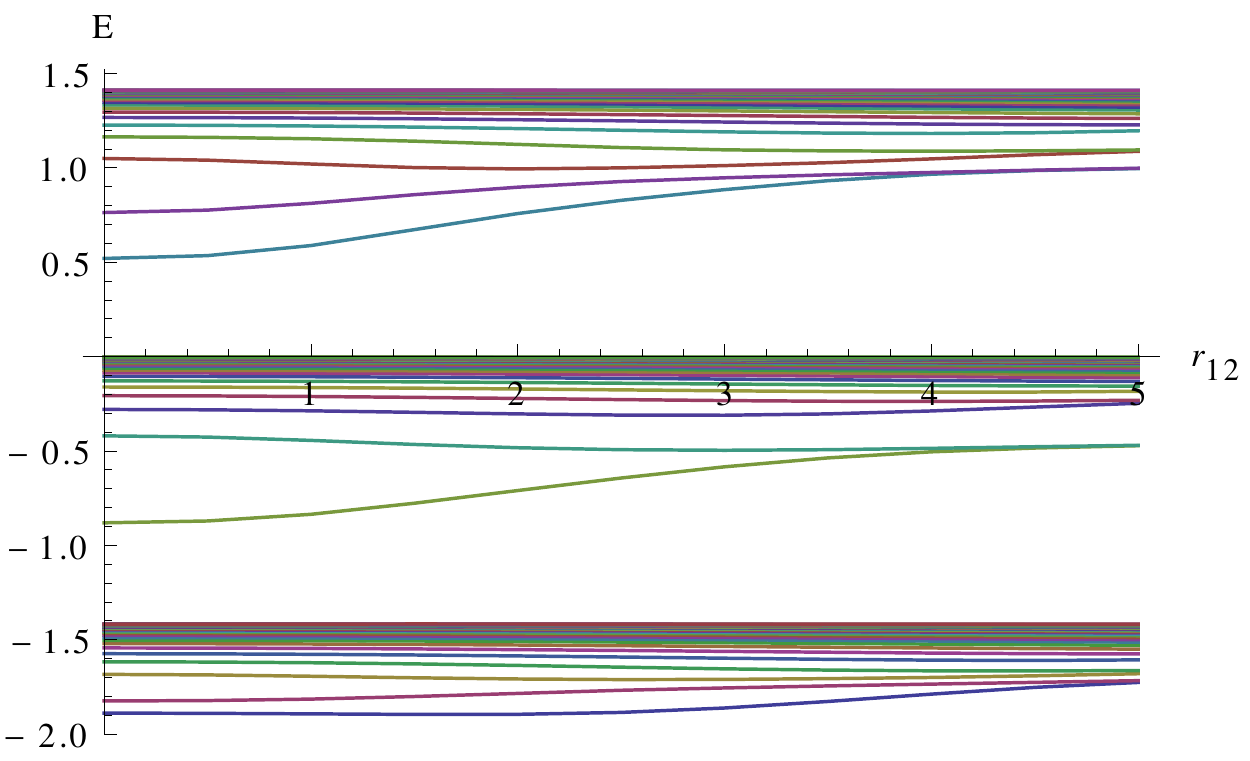}
\end{center}
\vspace{-0.6 cm}
\caption{\label{fig:energy levels}
Plot of 
1-particle energy sub-levels for several Landau levels as a function of distance between two positively charged Coulomb centers. 
Here 
$\alpha = 0.4$ and $d=0.05$.  A remarkable (but obvious) fact is that if all of the sublevels are populated, the distance-dependence 
of the total energy  disappears. }
\end{figure} 

 

\section{Graphene with charged impurities in magnetic field} \label{sec:1particle}
It is well-known that the one-particle energy levels of an ideal graphene in magnetic field $B$ are given by degenerate Landau
levels \cite{Geim2005,QHE2005,QHE2005Gusynin,GrapheneDisorder2006}:
\beq \label{LL}
  E_n = \sign(n) E_B \sqrt{2 |n|}\; , \; n \in \mathbb{Z}
\eeq
 where $v_F \approx 10^6 {\rm m/s}$,  
\beqa \label{lB}
l_B = \sqrt{\hbar/(e B)} \approx 26/\sqrt{B/({\rm Tesla})}\ {\rm nm} \\
\label{EB}
E_B \equiv \frac{\hbar v_F}{l_B} \approx 26 \sqrt{B 
/(\text{Tesla})} \meV.
\eeqa 
Each Landau level (LL) is degenerate with density 
$ 4 \frac{e}{2 \pi \hbar} B $ per unit area. Working with finite area, it is convenient to use the basis of Landau
wave-functions in polar coordinates. Defining an oscillator radial eigenfunction: 
\beqa
&g_{n,m}(r) \nn  =  \\ 
&\nn e^{-\frac{r^2}{4}} r^{\left| m\right| } \sqrt{\frac{2^{-\left| m\right| } (\left| m\right| +n)!}{2 \pi n! 
(\left| m\right| !)^2}} \,
   _1F_1\left(-n;\left| m\right| +1;\frac{r^2}{2}\right)
\eeqa
we have for the wave-functions (  $m < n$ , \; $n>0$ ):
\beqa  \label{basis}
&\psi_{0, m}(r,\phi) = \left(\bea{c} 0 \\
                              e^{i m \phi} g_{0,m}(r)  \ea  \right),  m \leq 0  \\
 \nn
&\psi_{\pm n, m}(r,\phi) = \frac{e^{i m \phi}}{\sqrt{2}} \left(\bea{c} \mp e^{-i \phi} g_{n-\frac{m-1+|m-1|}{2},m-1}(r) \\
                             i g_{n-\frac{m+|m|}{2},m}(r)  \ea  \right) 
\eeqa 
The solutions near the second $K$-point are obtained by acting with $i \sigma_2$ on the spinor:  $\(\ba{c} \Psi_1 \\ \Psi_2\ea \) \to 
\(\ba{c}\Psi_2 \\ -\Psi_1\ea \right)$. 

When the Coulomb impurity potential is present, the orbital (index $m$) degeneracy of Landau levels is lifted \cite{Kharitonov}.
This has been calculated\cite{Gamayun,Semiclassical, Kharitonov} and demonstrated experimentally \cite{Kharitonov} for one Coulomb 
impurity. Below we consider several impurities. 

\begin{figure*}[htb]  
$\begin{array}{ccc}
\hspace{-0.8 cm}\includegraphics[scale=0.4]{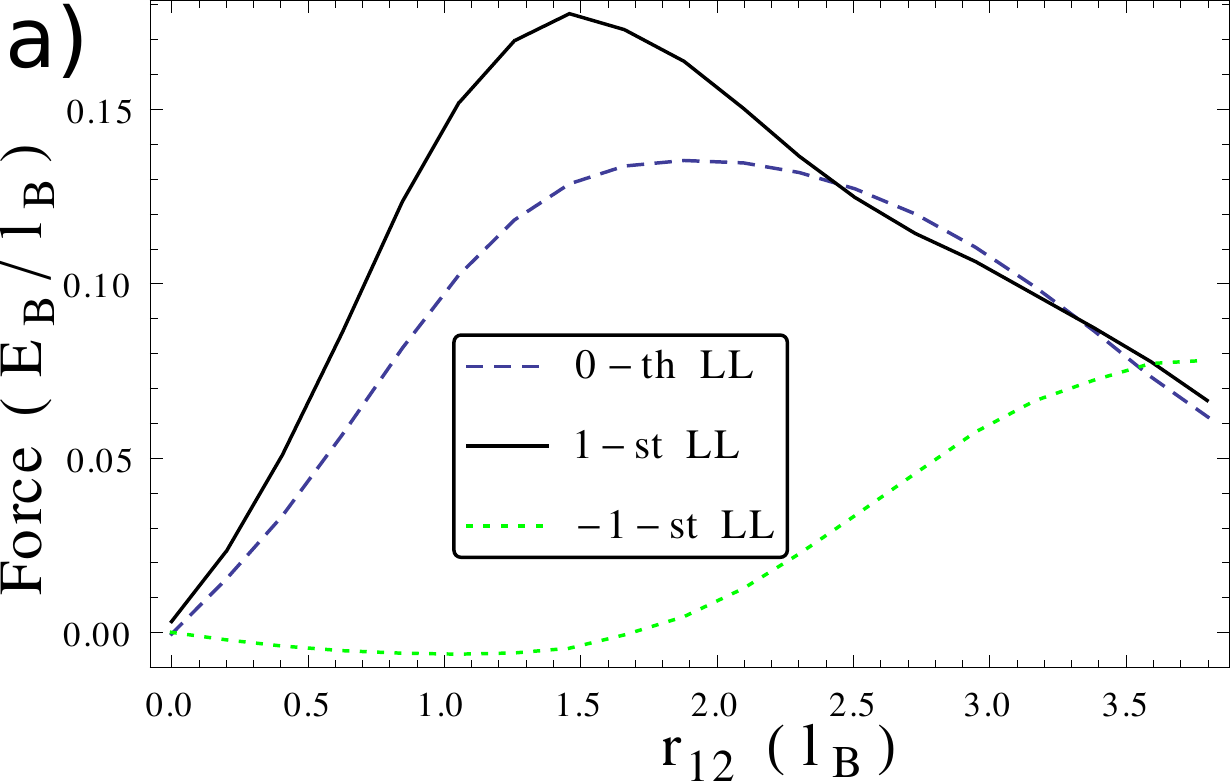} 
 \hspace{-0cm}&

\includegraphics[scale=0.4]{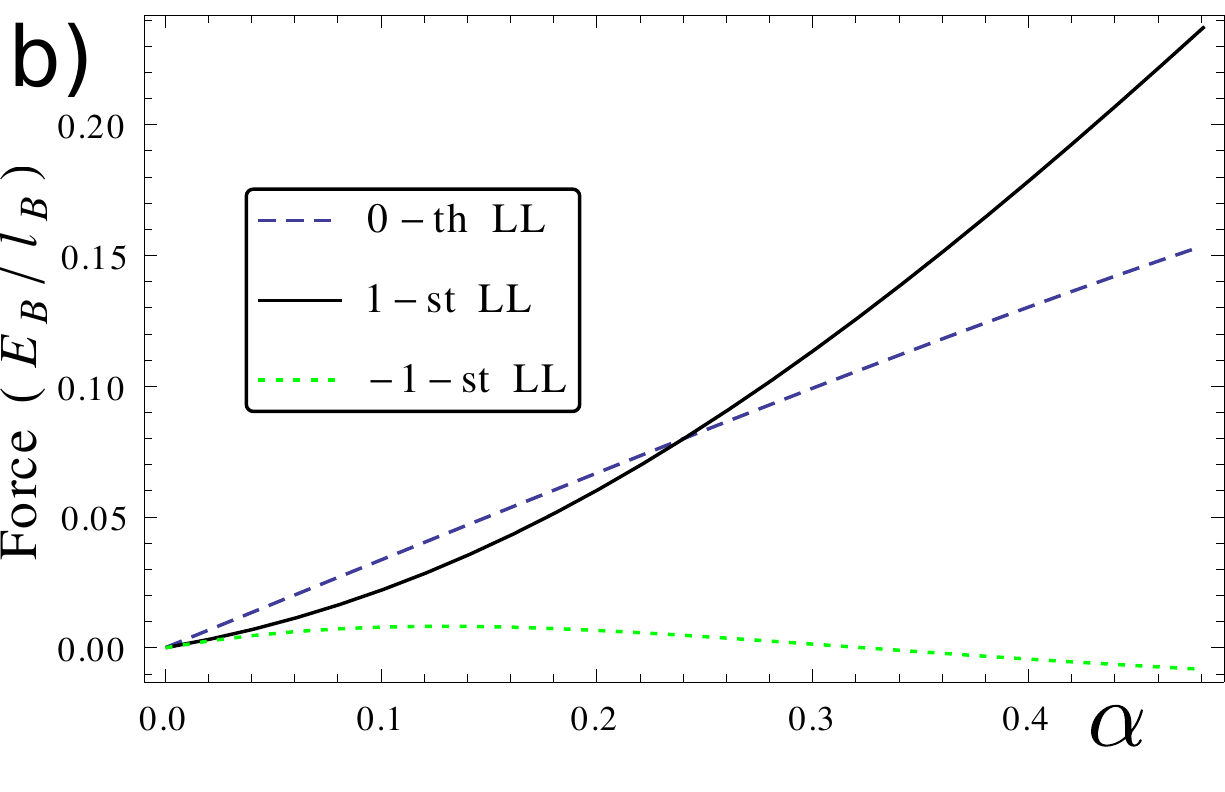}
\vspace{0.cm}
 & 
\includegraphics[scale=0.4]{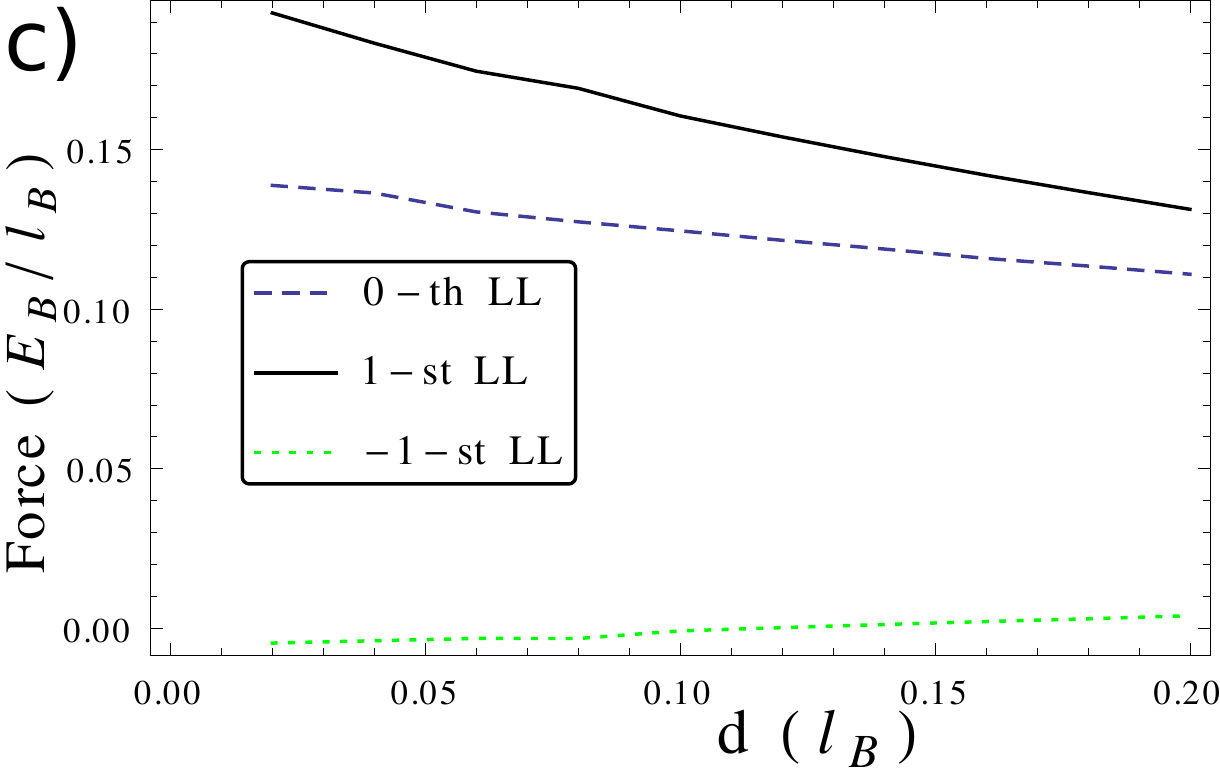}
\end{array}$
\caption{\label{fig:nonlinearity} \label{fig:ddependence} 
Binding force of two ions  bound by one electron: a) as a function of inter-ion distance $r_{12}$  for $\alpha=0.4$ and $d=0.05 l_B$;  b) 
as a function of coupling $\alpha$ for $d=0.05 l_B$ and $r_{12} = 1.5 l_B$;
c) as a function of $d$ (distance of ions from the graphene sheet) for $\alpha=0.4$, $r_{12} = 1.5 l_B$}
\vspace{-0.5cm}
\end{figure*}

Consider a superposition of Coulomb potentials of the form 
\beq
U(\vec r) =-\frac{e^2}{4 \pi \eps_0 
\eps} \sum_i \frac{1}{\sqrt{(\vec r-\vec r_i)^2+d^2}}
\eeq Here the parameter $d$ is a vertical displacement of impurity from the graphene sheet,  but 
it can also be used to model a finite localization length of impurity wave-function \footnote{This way we may consider a hole in the 
``dead layer'' of SiC epitaxial graphene as a mobile impurity.}. 
 Here, there are two physical cases that may be considered:
\begin{itemize}
 \item  The wave function of the localized impurity exceeds the graphene lattice scale,  $d \gg a$, but still $d \ll l_B$. This will be 
the main topic of the present work.  This
may be relevant to the case when localization of impurity is determined by Moire pattern formed by incommensurate substrate and 
graphene. For epitaxial graphene on Si-terminated SiC this scale is estimated as $d \sim 2 \,{\rm nm} $. 
Then the impurity potential is smooth on the lattice scale (note that it is still sharp on the scale of magnetic length) and 
single-valley continuum approximation works well,  we may neglect the inter-valley scattering. 
This means that the valley-mixing 
splitting is less than interaction-induced energy and multi-electron wave functions are determined by long-range Coulomb interactions.
This case is most universal since it depends only on the magnetic length scale and effective coupling constant. Additional universality 
comes from the fact that   
the dependence of binding force on the effective range $d$ of 
impurity wave function is weak,
see Fig.\ref{fig:ddependence}. 
 \item  If the impurity is more like a point charge, then the impurity potential is sharp and non-universal lattice-scale details come 
into play and cause the valley
mixing. Continuous approximation \eq{continuous} breaks down near the impurity center since sharp potentials cause inter-valley 
scattering. This is discussed in section \ref{sec:Valley Mixing}.
We discuss several lattice-related effects, but the main conclusions of the continuum-model considerations still hold true.
\end{itemize} 
  
The equations for single-electron energy levels in graphene in the magnetic field $B$ and any Coulomb potentials   can be 
rewritten 
\cite{Gamayun} in 
units of  magnetic length $l_B$, magnetic energy $E_B$ and  dimensionless coupling  
\beq
\alpha = e^2/(4 \pi \eps_0 \hbar v_F \eps_\text{eff}) = 2.19/\eps_\text{eff} 
\eeq 
with the effective dielectric constant $\eps_\text{eff} = (\eps_1+\eps_2)/2$ coming from substrates on both sides of graphene and from
graphene by itself.  For example, $\alpha \approx 
1$ on SiO$_2$ substrate \cite{Kharitonov} and $\alpha \approx 0.4$ on SiC.    
In dimensionless units the equation to solve is \cite{Gamayun}
\beq
 [\sigma_1 (i \d_x - y) - i \sigma_2 \d_y] \Psi =  \(E - \sum_{i=1}^N \frac{\alpha}{\sqrt{(\vec r-\vec r_i)^2+d^2}}\) \Psi
\label{continuous} 
\eeq
For ``molecules'' bound by more than one electron we will account for e-e interactions below. 
For performing the computations we evaluate the matrix elements of the impurity potential in the basis (\ref{basis}) truncated to 
several Landau levels (of the order of 10) and orbital states (of order of 30) and then do an exact diagonalization. 
In the case of smooth potentials it is sufficient to consider only one valley. The truncation 
of 
orbital states we use corresponds to a circular box truncation of space and to avoid unphysical boundary contributions we had to 
smoothly cut-off the Coulomb potential at large distances (of order of $4 l_B$). As a result, for inter-ion distances up to $2 l_B$, the 
precision is better than $1\%$ while for larger distances the error may get higher.

The Landau levels that are completely
filled do not contribute significantly to the energy of ions as a function of their separation (but they do contribute to renormalization
of chemical potential \cite{2DEGEnergy, 2DEGSubstrateEnergy}). 
This has been verified numerically and
is seen analytically in the
leading order of perturbation theory in the potential: 
\beqa
 E(r_{12}) &\sim&  \sum_m \< \psi_{n,m}^{(0)} |V(\vec r - \vec r_1) +V(\vec r - \vec r_2)   | \psi_{n,m}^{(0)}\>  \nn \\
&=& 2 \sum_m \< 
\psi_{n,m}^{(0)} |V(\vec r)| \psi_{n,m}^{(0)}\>  = const  
\eeqa  
where we used that the full degenerate set of wave-functions, corresponding to a given Landau level (enumerated by $m$) maps 
to 
itself under translations (up to a unitary transformation), and so the above sum is independent of the impurity positions.  

The situation changes drastically if only one or several lowest Landau sub-levels are filled with electrons. 
This happens when graphene is in Quantum Hall plateau state corresponding to the chemical potential being in the gap below the 
band edge (i.e. between the unperturbed  
LLs in 1-particle picture). 


  Let us start with two positive charges $N=2$.  The single-particle
energy levels of electrons in graphene are presented on Fig.\ref{fig:energy levels}. 
Note that only the lowest sub-levels are fully meaningful as filling more levels requires to account for the e-e interactions.

When the distance between the two
ions is of the order of magnetic length $l_B$, the lowest energy electron wave-function is {\it centered  in the middle between the ions} 
and 
plays the role of 
hybridization cloud that binds them, Fig. \ref{fig:Wave Function}. When only this lowest-energy state is filled,  the strong 
dependence of energy on the distance $r_{12}$ between ions appears, Fig. \ref{fig:energy levels}, creating the attractive force, Fig. 
\ref{fig:nonlinearity}.  If the distance between charges exceeds roughly $3 l_B$,  the lowest energy electron wave-function becomes 
centered near each of the individual charges and its energy depends on $r_{12}$ as $E \sim - 1/r_{12}$.

In the leading order of perturbation theory, the binding force is proportional to $\alpha$. 
Notably, the mutual Coulomb repulsion of ions is also proportional to $\alpha$:
\beq
  E_{\text{Coulomb}} = \frac{\alpha}{r_{12}}
\eeq
Thus, in the leading approximation the distance
where the attraction would balance the repulsion is independent of $\alpha$.  We stress here that when expressed in magnetic length 
and energy units, there are essentially no free 
parameters in the 
problem and the Coulomb repulsion of ions is of the same order of magnitude as hybridization attraction. \footnote{This will not be the 
case for the ordinary 2D electron gas since the quasiparticle mass will essentially enter the game.} 
It is a-priori not at all 
clear if stable bound states of ions can form. Moreover, strong bound states form only when the chemical potential $\mu$ is near the 
0-th and 1-st Landau levels. When $\mu$ is near the other Landau levels (see Fig. \ref{fig:energy levels}), the 
$r_{12}$ dependence of electron energy is substantially weaker leading to weaker binding of ions. The main problem with bound 
states in the higher plateau states is that they will not survive graphene rippling and temperature, 
thus we concentrate on the most robust $\nu \approx  \pm 2$ plateau states below.      
 
Let us first study the dependence on the coupling constant $\alpha$, see Fig. 
\ref{fig:nonlinearity}b.  We observe that 
the 0-th LL is almost protected from non-linearity, while a significant non-linearity appears in the higher LL already at $\alpha 
\sim 0.1$. 
The 1-st and -1-st levels must have the same energies in the leading order of perturbation 
theory 
as their wave-functions differ only by a relative sign of sublattice components,  but we see that  these levels
split already from $\alpha \sim 0.1$. This could be expected due to large Coulomb field near an individual impurity.
Similar asymmetry is present in the results of ref.\onlinecite{Kharitonov}.
The dependence on the distance $d$ of the ion from graphene (or, on the localization length of charge wave-function) is weak when this 
distance is much less than magnetic length, see Fig.\ref{fig:nonlinearity}c.
   
Below we present a particular example of $\alpha=0.4$, $d=0.05 l_B$ ,  the results are universal for 0-th LL, while for the 1-st LL 
the binding is seen to grow a bit faster than linear in $\alpha$.  

\begin{figure*}[htb]  
$\begin{array}{ccc}
\hspace{-0.8 cm}\includegraphics[scale=0.4]{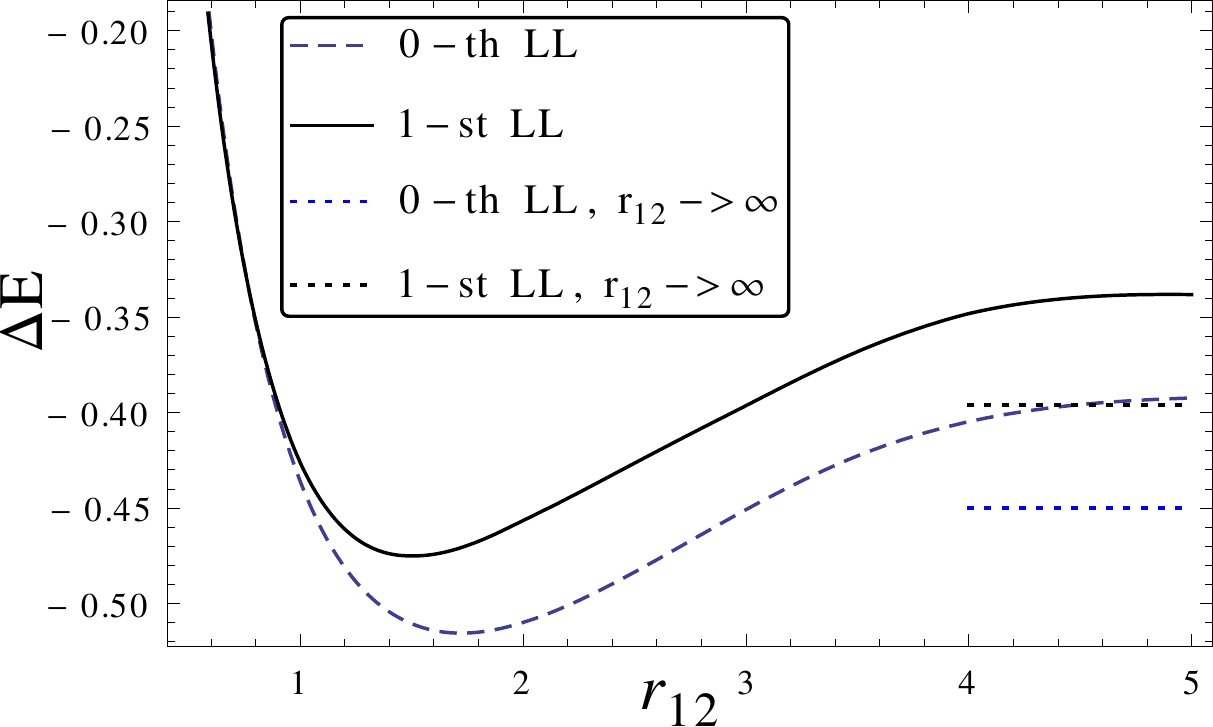} 
 \hspace{-0cm}&

\includegraphics[scale=0.4]{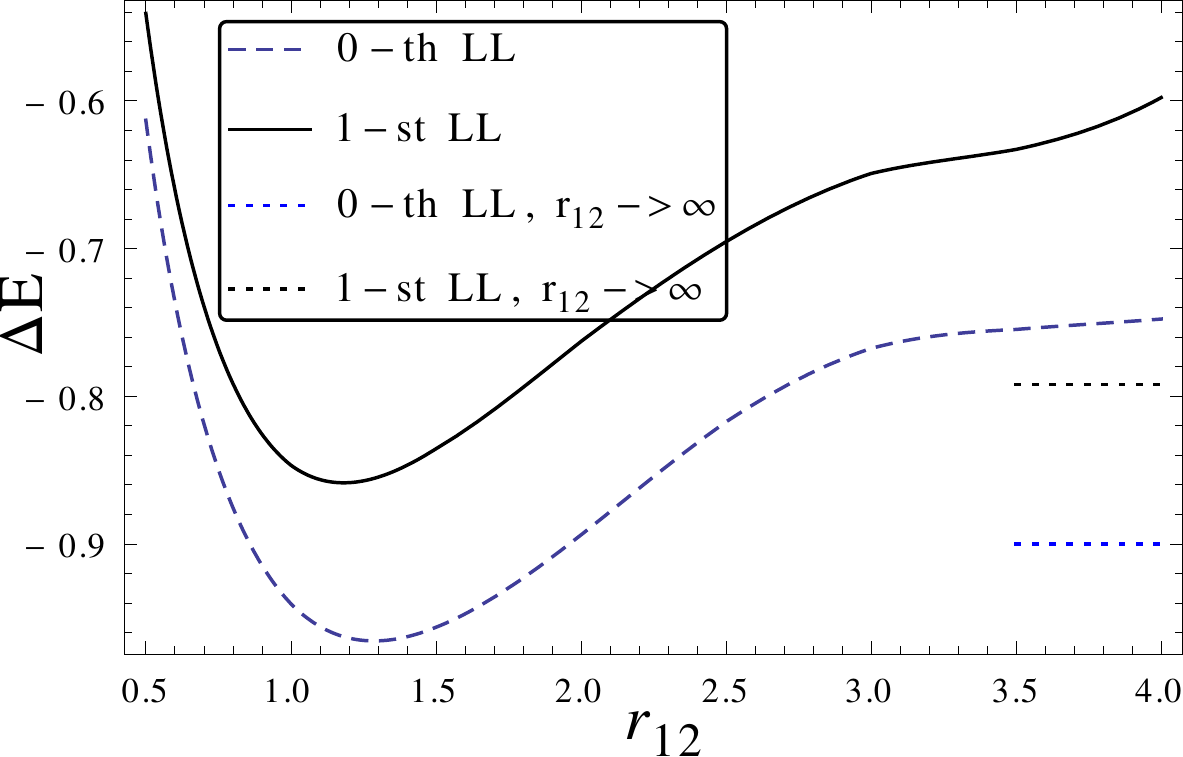}
\vspace{0.cm}
 & 
\includegraphics[scale=0.4]{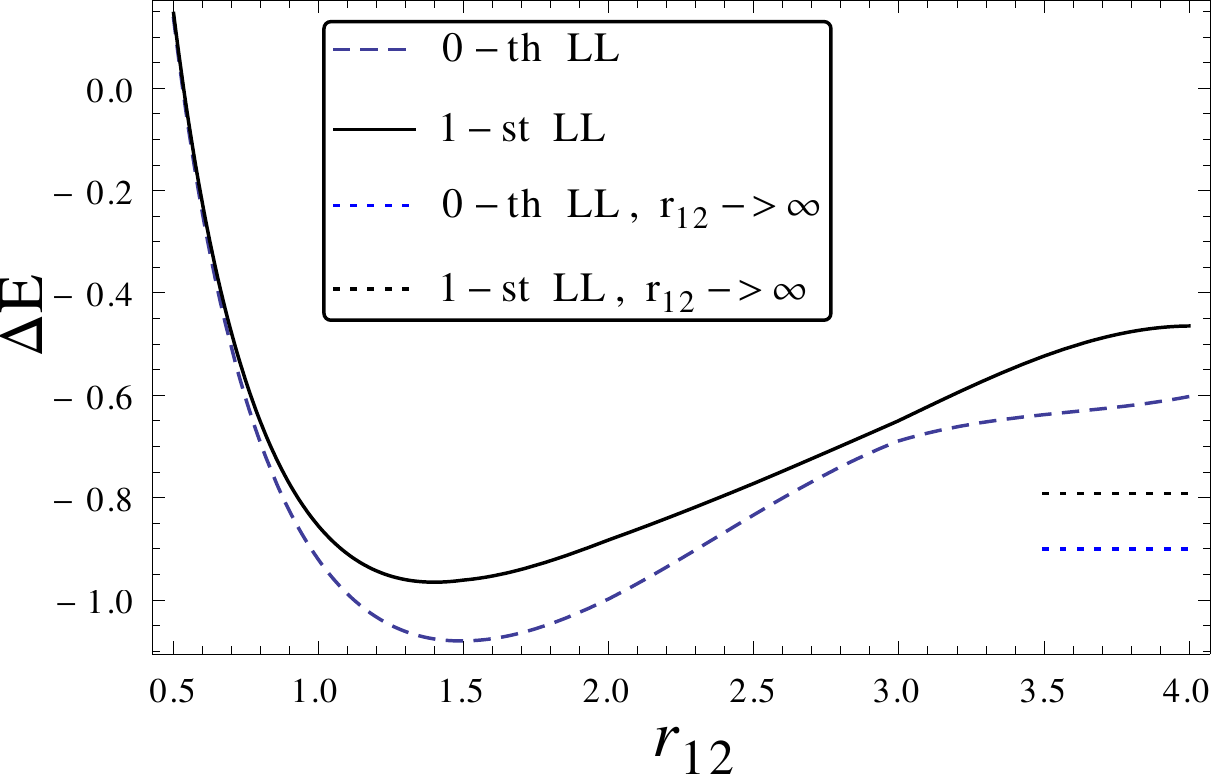}
\end{array}$
\caption{\label{fig:2imp1e}   \label{fig:2imp2e}  \label{fig:3imp2e}  
Energy (electron energy + Coulomb repulsion of ions) of stable macro-molecules. Electron energy is counted from Landau level edge (fixed 
to free-particle values $0$ and 
$\sqrt{2}$ for 0-th and 1-st LLs respectively). Here $\alpha = 0.4$ and $d=0.05 l_B$. Dotted lines mark the $r\to \infty$ asymptotic to 
show stability.  a) Two ions bound by one electron in graphene; b) Two ions bound by two electrons in symmetric orbital state; c) Three 
ions forming equilateral triangle bound by two electrons in symmetric state
}
\vspace{-0.5cm}
\end{figure*} 

The results for two positive ions bound by one electron are presented on Fig.\ref{fig:2imp1e}a. We observe an absolute minimum in 
the energy at a distance $r_{12 \text{min}} \approx 1.5 l_B$ near the 1-st LL 
or $1.7 l_B$ near the 0-th LL. 

If the chemical potential decreases below a critical value, determined as the energy per electron in the bound state, see Table 
\ref{fig:phase diagram},  there will be no bound states for positive 
ions, but, at some point, we start getting bound states for negatively charged
impurities bound by holes. The calculations and results for negatively charged impurities are exactly the same and obtained by 
going to hole picture. 
\comment{
\begin{figure}[t]  
\begin{center}
\includegraphics[scale=0.5]{1imp.pdf}
\end{center}
\vspace{-0.9 cm}
\caption{\label{fig:1imp}  Binding energy of electron in the 
field of one Coulomb impurity as a function of dimensionless coupling $\alpha$. 
Curves for Landau levels -1, 0  and 1 are given, energies without impurity ($-\sqrt{2}$, $0$ and $\sqrt{2}$) are subtracted. 
}
\end{figure}

\begin{figure}[t]  
\begin{center}
\includegraphics[scale=0.5]{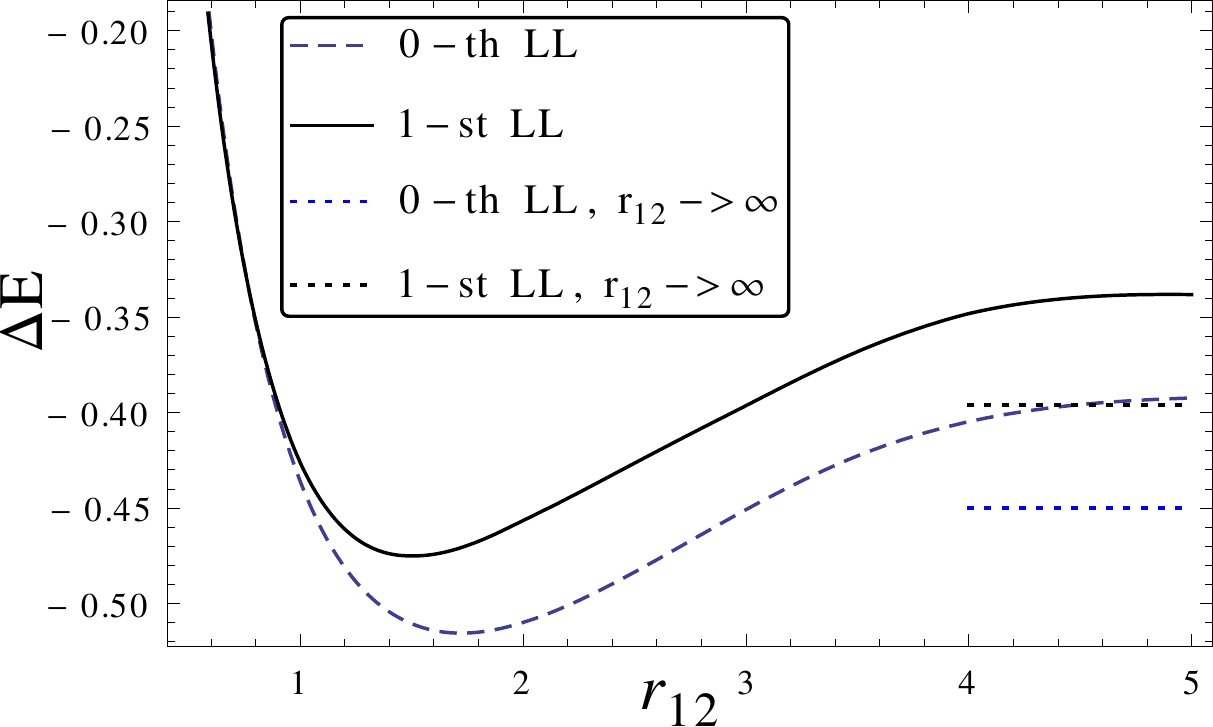}
\end{center}
\vspace{-0.6 cm}
\caption{\label{fig:2imp1e} 
Binding energy of a stable macro-molecule, consisting of two positively-charged ions bound by one electron in graphene. 
The two curves correspond to the filling fraction being near +2 (filling of lowest sublevel of 1-st LL) or near -2  (0-th LL) 
respectively. Here $\alpha = 0.4$ and $d=0.05 l_B$. Dotted lines mark the energy of infinitely separated impurities (one is screened with 
an electron, another is single),
}
\end{figure} 
}
 
Analogous calculation 
shows 
that the configurations with three symmetrically positioned impurities bound by one electron is unstable: despite an appearance of local 
energy minimum, 
the configuration would gain energy if deformed to a bound pair with the third ion repelled to infinity. 

\section{Multi-electron states} \label{sec:multi-electron}
Now consider multi-electron bound states.  The two-electron bound state can be 
in a symmetric 
or antisymmetric orbital state. In a conventional molecule this would correspond to spin singlet and triplet respectively,  but in 
graphene one has an additional valley degeneracy \cite{ReviewMagnetic,Review} and the full $SU(4)$ symmetry is approximately 
respected (neglecting the Zeeman splitting and valley-mixing).
Group theory tells us that $\mathbf{4} \otimes \mathbf{4} = \mathbf{10} \oplus \mathbf{6}$ in spin-valley space,
so, there are 10 possibilities to form antisymmetric orbital state and 6 for symmetric one.  If valley-mixing effects discussed in 
section \ref{sec:Valley Mixing} are important (exceed the temperature scale), then orbital-symmetric state will be the usual 
spin-singlet. 
     
To find a reasonable
approximation to the energy, we use the variational Hartree-Fock method with the basis formed by Slater determinants of low-energy 
1-particle eigenstates found above. 

Looking at Fig.\ref{fig:energy levels}, 
we note that near the 1-st LL the two lowest single-particle energy sublevels grow with $r_{12}$  thus can 
potentially bind the ions and can participate in antisymmetric orbital wave-function. For the 0-th LL only the lowest sublevel binds the 
ions while electrons in the higher levels do not tend to hybridize, thus the orbital wave-function should be symmetric for a stable 
ion-binding. These conclusions were checked by explicit computations and comparison of the energy. 

Consider two electrons in the field of a single ion. For $\alpha=0.4$ we have energies  $ -0.28 $ , $-0.21$ per electron in 
0-th and 1-st LL respectively for symmetric state   and  $ -0.24 $ , $-0.22 $  respectively for antisymmetric orbital. Thus we expect 
symmetric state near 0-th LL and antisymmetric state near the 1-st LL.  Calculation shows that one unit-charge ion cannot hold more than 
2 
electrons in the lowest-energy states: e-e interactions make this too expensive.  

Now we consider two positive charges and two electrons. Calculation shows that symmetric orbital state is preferred.   
The resulting hydrogen-like molecule is very stable with optimal inter-atomic distance
$1.2 l_B$ and $1.3 l_B$ for 1-st and 0-th LL,  Fig. \ref{fig:2imp2e}b.

The same calculation can be repeated for 3 ions bound by two electrons, see Fig.\ref{fig:3imp2e}c. We compare the triangle configuration
of ions with a linear chain geometry and find that equilateral triangle geometry has lower energy. The described equilateral triangle with 
2 electrons is prominent for providing the lowest possible energy per electron, thus, 
it is this configuration that appears first in the phase diagram, Table. \ref{fig:phase diagram}.
 Four ions cannot be bound by two 
electrons. 

Considering now 3-electron states in Hartree approximation we found that three electrons cannot form a one-centered 
wave-function to bind any number of ions since the e-e interaction gets too high and the resulting energy gain can by no means compete 
with the energy of far-separated smaller clusters described above.  The situation is different from the ordinary molecules since the shape
of wave-function is mostly determined by Landau level number and not by Coulomb potential of charges. For the lowest Landau levels 
that are most robust, the shape of wave-functions is essentially circular (similar to s-state). The above considerations still do not 
exclude the possibility of larger multi-electron bound states with {\it multi-centered electron wave function}, but quantitative study
of such configurations is beyond the scope of this paper. 

\begin{table}[h]
\caption{\label{fig:phase diagram} Phase diagram for $N$ positively charged ions bound by $n$ electrons at $\alpha=0.4$, 
$d=0.05 l_B$;  
for negatively-charged ions 
one has to replace $\mu \to -\mu$. $\mu_{min}$ gives the minimal chemical potential for existence of a given phase.  Chemical potentials 
should be understood as taken relative to Landau 
level band edges, which are here renormalized to free-particle values $0$ and $\sqrt{2}$ for easy match with free-particle levels. 
$r_{ij}$ is the optimal distance between ions, i.e. the size of the molecule, measured in units of magnetic length $l_B$. 
$E$  
is a minimal possible (binding) energy per molecule, which is achieved when $\mu$ tends to the band edge ($\mu=0$ for configurations 1 
\ldots 5 and $\mu = \sqrt{2}$ for configurations 6 \ldots 10). For lower $\mu$ the energy to consider is $E - n \Delta \mu$.  The last 
column, $\mu$ interval, indicate 
the range of chemical potentials where the phase gives the minimal energy per charge (see \eq{EPerIon}). }
\begin{tabular}{|c|c|c|c|c|c|c|}
\hline 
\textnumero & $ \mu_{\text{min}}$  & $N$ ions  & $n$ electrons & $ r_{ij}$ & $E$ & $\mu$ interval \\ \hhline{|=|=|=|=|=|=|=|}
1 & -0.93 & 3 $\triangle$ & 2 symm. & 1.5  & -1.06 & \parbox[c]{1.7cm} {[-0.93,-0.75] \\ {[-0.55,-0.38]}}  \\ \hline
2 & -0.75 & 2 & 1 & 1.7  & -0.52  & [-0.75, -0.55]  \\ \hline
3 & -0.63 & 2 & 2 symm. & 1.3  & -0.96&  [-0.38 ,-0.08]  \\ \hline
4 & -0.45 & 1 & 1 & -    & -0.45 &  $\emptyset$ \\ \hline
5 & -0.28 & 1 & 2 symm. & -    & -0.56 & [-0.08,0] \\ \hline
6 &  0.51 & 3 $\triangle$ & 2 symm. & 1.4  &  -0.95 
                                                    & \parbox[c]{1.7cm}{[0.51,0.67] \\ {[0.95,1.09]} } \\ \hline
7 &  0.67 & 2 & 1 & 1.5  &  -0.47  & [0.67,0.95] \\ \hline
8 &  0.82 & 2 & 2 symm. & 1.2  &  -0.85 & $[1.09,1.39]$ \\ \hline
9 &  1.02 & 1 & 1 & -    &  -0.39 &  $\emptyset$ \\ \hline
10&  1.19 & 1 & 2 anti-s. & -    &  -0.45 &  [1.39,1.41] \\ \hline
\end{tabular}
\end{table}

\section{Phase diagram} \label{sec:phase diagram}
Having studied the simple macro-molecules separately, the results may be combined in a phase diagram, see Table \ref{fig:phase 
diagram}.  
In each of the above states one can find the energy per electron that bind the molecule and thus find a minimal electron 
chemical potential $\mu_{\text{min}}$ for such molecule to appear. 
For a given electron chemical potential, several molecule configurations may be possible.  Let us assume that the number of ions
and electron chemical potential are fixed. For a molecule with $N$ ions bound by $n$ electrons  we compute the free energy gain per ion 
by the formula:
\beq \label{EPerIon}
 \Delta E_{\text{ion}} = \frac{ (E_{\text{binding}}+E_{\text{Coulomb}})- \mu n}{N} =\frac{E-\mu n}{N} 
\eeq   
The configurations with minimal $\Delta E_{\text{ion}}$ will deliver the minimal free energy at zero temperature, see Fig. 
\ref{fig:PhaseDiagram}. In particular, Fig.\ref{fig:PhaseDiagram} shows that the state  of one electron bound by one ion (lines 4 and 9 
on the plot) is never the lowest-energy state.  At high chemical potentials (close to the vacuum LL) the state with 2 electrons per one 
ion  wins  (lines 5 and 10). All the other states correspond to bound states of 2 or 3 ions, which are realized for $\mu \in 
[-0.93,-0.08] \cup [0.51, 1.39]$  for the example $\alpha=0.4$ considered.   

Changing $\alpha$ in the leading approximation just linearly scales the phase diagram around the vacuum Landau 
level (0 and 
$\sqrt{2}$). As is seen from Fig. \ref{fig:nonlinearity}, the binding of molecules near $\nu = 2$ should additionally increase with 
increasing $\alpha$ due to noticeable non-linearity.   

To come to physical conclusions,  consider a realistic example of randomly positioned charges.
For example, consider 4 charges and the $\nu \approx 2$ plateau state.  Looking at the Table \ref{fig:phase diagram} we can find the 
states that deliver the best binding energy per charge: 
\begin{itemize}
\item if
the chemical potential is $\mu \in[0.51,0.67]$, 3 charges may be bound with 2 electrons in the triangle (state \textnumero 6) and the 
fourth is repelled to infinity since no electrons can be bound to it. In reality, it would be hard to create such state since one needs
to surpass a significant Coulomb repulsion gap to form the triatomic molecule.
\item if $\mu \in[0.67,1.02]$,  two diatomic ions (two ions bound with one electron) would be the preferred state (again, it will be hard 
to create). The state 1 will continue to exist as well.
\item for $\mu>1.02$,  a single charge can bind electron, forming a neutral combination. Now it is much easier to form molecules since 
there is no long-range Coulomb barrier. In particular, the triatomic configuration (\textnumero 6 in the table) is now much easier to 
form, note that this configuration has unit total positive charge. One remaining charge will have one electron bound to it and this 
neutral combination will be attracted by the induced dipole moment to any charged object. Thus, a 4-atomic molecule may 
be formed. The precise determination of its energy is beyond the scope of this paper.
\item
 for $\mu>1.09$ the diatomic charged configurations (\textnumero 6) would shrink in size (from 1.5 to 1.2) to form the neutral diatomic
molecules (\textnumero 8).  
\end{itemize}

Analyzing the above example we may draw the following conclusions. If there are randomly-positioned same-charge mobile impurities,
the active recombination process would start only when  neutral states of one impurity and one electron (\textnumero 4 or \textnumero 9) 
may be formed. For lower chemical potential the stable bound states exist but they are hard to form due to significant Coulomb 
repulsion barriers. The most relevant recombination channel is to form neutral hydrogen-like molecule (\textnumero 3 or \textnumero 8).
In general, the clustering process is mediated by the induced dipole attraction of neutral and charged macro-molecules.

If we are dealing not with the real ionic impurities, but with the holes dynamically appearing in the surface layer of substrate below 
graphene,
the Coulomb barriers for formation of bound states are much lower and all the states in Table \ref{fig:phase diagram} are relevant. Their 
appearance will be governed by the {\it local} chemical potential formed by other non-mobile charges.  

It should be kept in mind that all these calculations make sense when concentration of mobile impurities is low: $n_{imp} \ll 
l_B^{-2}$. Correspondingly, the filling factor of electrons that bind these impurities must be very close to the perfect values $\pm 2$.
These conditions could easily be achieved {\it locally} in the incompressible regions appearing dynamically in the QHE physics. All the 
``extra'' electrons are expelled out to compressible regions.    



\comment{
\begin{figure}[t]  
\begin{center}
\includegraphics[scale=0.5]{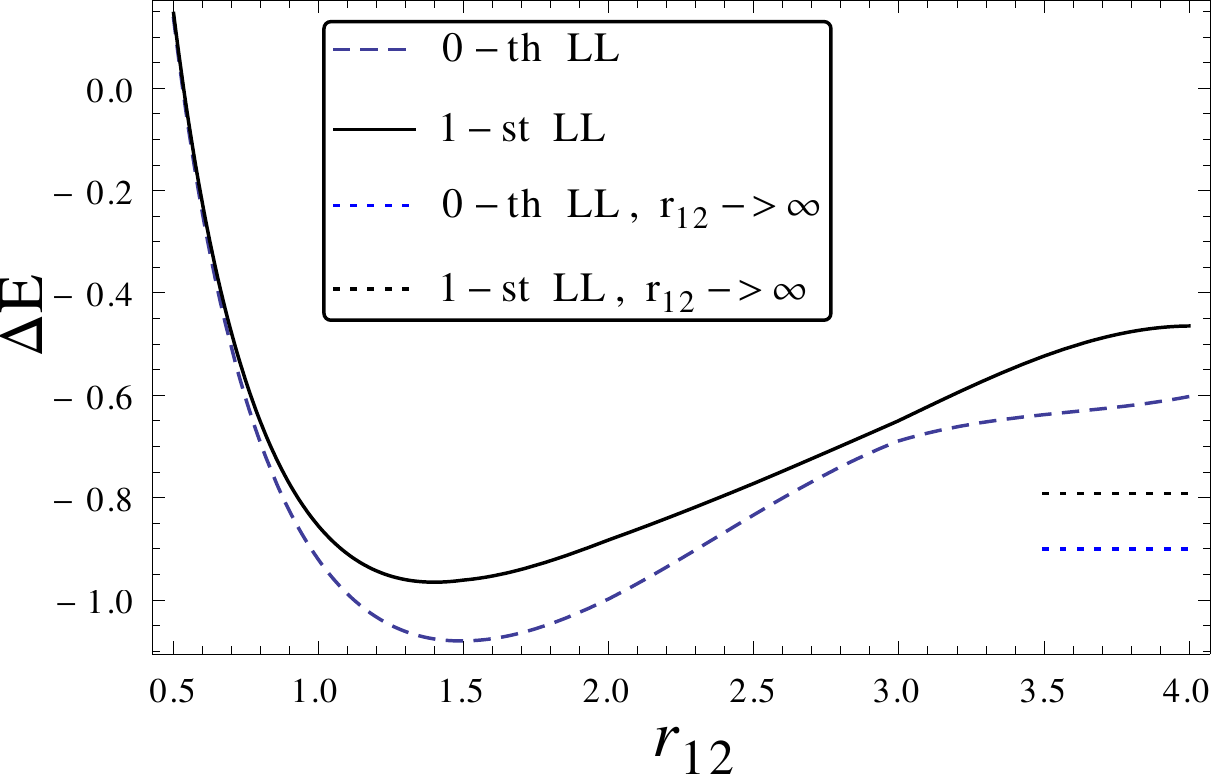}
\end{center}
\vspace{-0.6 cm}
\caption{\label{fig:3imp2e} 
Energy of three positively-charged ions in triangle geometry bound by two electrons in graphene, 
$\alpha=0.4$, symmetric state.   Dotted lines indicate the total electron binding energy for all impurities taken apart.}
\end{figure} 
}

\section{Valley mixing and lattice effects} \label{sec:Valley Mixing}
In case of sharp potentials we need to consider the doubled basis, involving the Landau level states in both valleys. 
Naively, the matrix element between Landau states could be computed by discrete lattice summation in the circle around the 
impurity and with continuum integration in the other regions. On the other hand, using only Coulomb interaction and Dirac bands in the 
immediate vicinity of the impurity  would be an oversimplification and the full (e.g.DFT) study including high-energy band
is needed. Thus, we 
follow a phenomenological approach and model
the impurity as smoothed Coulomb potential (as studied above) supplemented with point-like scatterers  sitting on A and (or) B 
sublattices.     
According to ref.\onlinecite{Wehling2009}, positively charged impurities (Li, Na,K, Cs) tend to be located in the center of graphene 
hexagons 
(``h-point'') and the negative impurities (Cl,Br,I) prefer to stay on top of carbon atoms (``t-point''). Thus, it is natural to expect 
sublattice-symmetric effective  potential for positive ions ($V_{Ai} = V_{Bi}$)  and sublattice-asymmetric potential for negative 
impurities. In the latter 
case, similarly to ref. \onlinecite{AbaninClustering08},  it is interesting to study the difference of same-sublattice and 
different-sublattice 
locations of impurities.  A general point potential is parameterized by its sublattice components $V_{AAi} , V_{BBi}, V_{ABi}$ at 
each of the impurity positions $x_i$. These components are naturally combined in $2 \times 2$ matrix $V(x_i)$.  

\begin{figure}[t]  
\begin{center}
\includegraphics[scale=0.5]{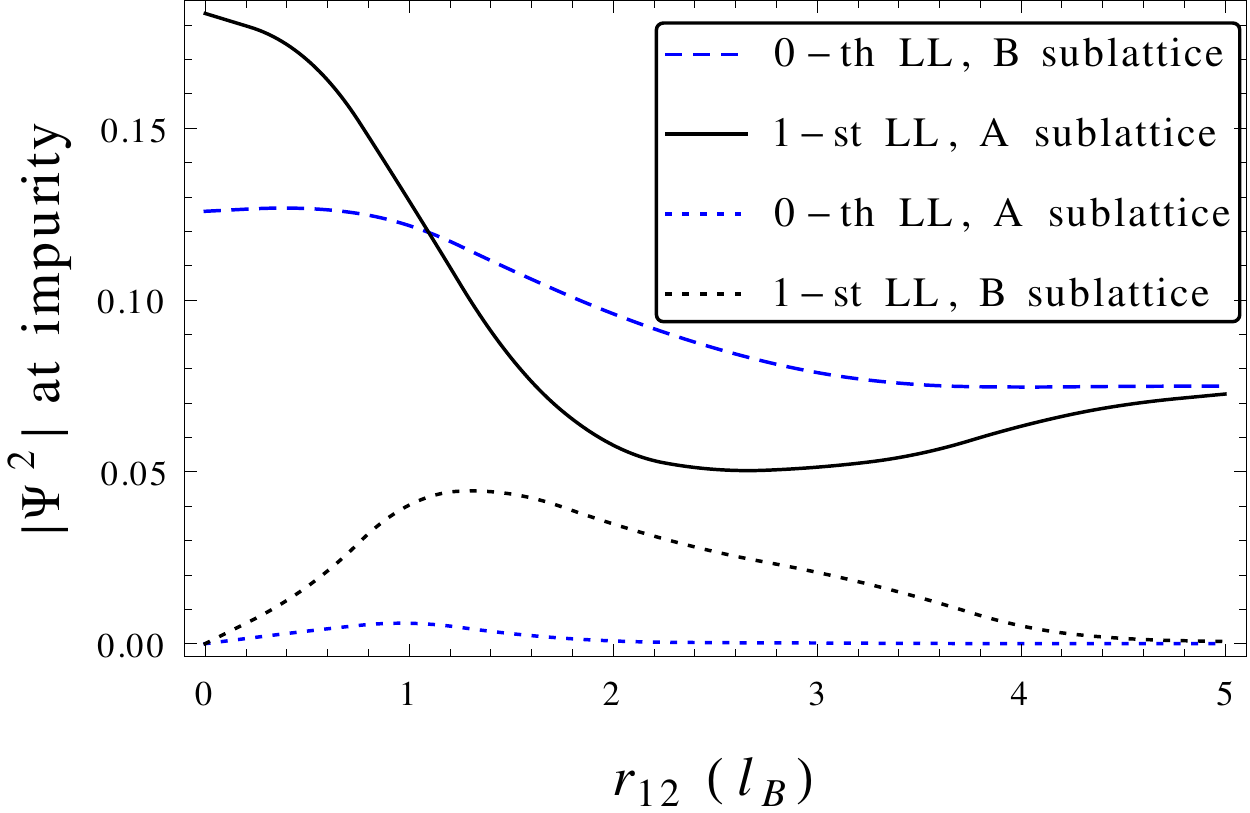}
\end{center}
\vspace{-0.6 cm}
\caption{\label{fig:Valley Mixing} 
Plot of sublattice components of wave-functions for lowest-energy sublevels of 0-th and 1-st Landau Levels  in the field of two 
charges at distance $r_{12}$.  Wave-functions are evaluated at the location of charged impurities.
}
\end{figure}

 In the leading order of perturbation theory, these potentials shift and split the single-valley energy levels discussed above.
Consider corrections to energies of 1-electron bound states in the field of $N$ impurities. The wave-functions with the same 
energy $E_0$ in the
second valley are obtained as $\Psi_{K'} = i \sigma_2 \Psi_{K}$. To discuss the lifting of this degeneracy in the leading order of 
perturbation theory we have the Hamiltonian 
$H = \left ( \ba{cc} E_0 + V_{11} ,& V_{12} \\
                      V_{12}^* , & E_0 + V_{22} 
              \ea \right) $   
with matrix elements
\beqa
  V_{11} &=& \sum_{j=1}^N \Psi^\dag(x_j) V(x_j) \Psi(x_j)  \\
  V_{12} &=& \sum_i \Psi^\dag(x_j) V(x_j) i \sigma_2 \Psi (x_j) \\  
  V_{22} &=& \sum_{j=1}^N \Psi^\dag(x_j) \sigma_2 V(x_j) \sigma_2 \Psi(x_j)  
\eeqa
The energy eigenvalues are:
\beq
E=E_0 + \frac{V_{11}+V_{22}}{2} \pm \sqrt{\(\frac{V_{11}-V_{22}}{2}\)^2+|V_{12}|^2}
\eeq
Note that the sublattice components of the wave-functions for the zeroth and the first Landau-levels are strongly asymmetric: one 
sublattice component is dominant for a chosen valley, see Fig.\ref{fig:Valley Mixing}.  
Hence, with a good accuracy, valley matrix elements are proportional to sublattice matrix elements: $V_{11} \sim \sum_i V_{BBi} 
|\Psi(x_i)|^2$, $V_{22} \sim \sum_i V_{AAi} 
|\Psi(x_i)|^2$ and $V_{12} \sim \sum_i V_{ABi} 
|\Psi(x_i)|^2$. For negatively-charged ions situated on the same sublattice, we get
the main non-zero matrix element $V_{11} = 2 V$. This shifts down the energy in one of the valleys by $2 |V|$.  If impurities are on 
the different sublattices we get $V_{11} = V_{22} = V$, which gives an equal twice smaller energy shift $|V|$ for both valleys. Thus, 
location of impurities on the same sublattice is energetically preferred and the corresponding electronic wave-function 
prefers this sublattice (assuming that $V<0$ 
).  

For positively-charged ionic impurities, one expects $V_{11} = V_{22}$ due to preferred location of impurity in the centers of  
hexagons. Then the sublattice-mixing matrix element  $V_{12}$ plays the leading role in lifting the valley degeneracy. 
The preferable wave-function
would then be a symmetric or anti-symmetric combination of both valleys or sublattices. 

Both effects advocated above lead to an extra gain in binding energy that would depend on the distance between the impurities, thus,
producing an extra force. The distance dependence is governed  by $|\Psi(x_i)^2|$, which, as already stated, is dominated by only 
one sublattice component, see Fig. \ref{fig:Valley Mixing}. 

Figure \ref{fig:Valley Mixing} illustrates that the effects related to scattering on sharp potentials do further increase the binding 
force. The 2-electron states discussed before
essentially fill the same orbital state with two electrons, hence, the discussed correction just doubles.     
The matrix elements 
discussed above scale with magnetic field as $|\psi^2(x_i)| \sim 1/l_B^2 \sim B$. The 
resulting extra contribution to energy scales as $B$ and the force scales
as $B^{3/2}$. This is to be compared with $B^{1/2}$ scaling of the main term in the energy.

\section{Discussion and conclusions} \label{sec:conclusions}
To conclude, we have shown that for a significant range of chemical potential values inside the gap between Landau levels  the charged 
impurities (or donor states in the substrate surface) can form stable bound states. The optimal distance between charges in the bound 
state is of the order of magnetic length $l_B=\sqrt{\hbar/(e B)}$. The binding energy scales as $E_B = \frac{\hbar v_F}{l_B}$ and
the binding force at optimal distance scales as $F \sim E_B/l_B \sim  v_F e B$.     

The above results were obtained for an ideal monolayer graphene sheet at zero temperature.  The realistic graphene may have other
non-mobile impurities, ripples, corrugations  and finite temperature. Clearly, mobile charge impurities can equally well form bound 
states with non-mobile ones.  Simultaneously, charged impurities may introduce smooth inhomogeneities in the chemical potential 
leading to replacement of chemical potential $\mu$ with a local chemical potential
$\mu + U_\text{impurities}$ in our considerations. It is also important to note that sufficient amount of mobile charged impurities 
leads to screening of potential landscape thus making it flatter on large scales. This may be one of the keys to understanding of 
exceptionally precise Hall quantization in epitaxial graphene\cite{TzalenchukNature}.   
 
 The temperature and short-range impurities lead to level-broadening.  As is clear from  Fig. 
\ref{fig:energy levels}, the binding appears when the lowest Landau sub-level 
is filled, while the next ones are empty.  The splitting between these levels is of the order of $\alpha E_B 
\approx 300 \alpha \sqrt{B/(\text{Tesla})}\, {\rm K} $  (see Fig. \ref{fig:energy levels}). The splitting of levels is twice smaller near 
the 1-st LL, but in this case the second smallest LL is also attractive and population of this level does not spoil the binding. 
So, with $\alpha \approx 0.4$  our results must survive the room temperatures and small amount of short-range impurities for magnetic 
fields above 10 Tesla  and even higher temperatures at larger fields. These conclusions are also supported by the experiment of ref. 
[\onlinecite{Kharitonov}].  
Note that at room temperature (and even below) many types of ionic impurities are mobile\cite{Wehling2009}. 

Another effect of finite temperature is an entropic contribution coming from the approximate spin and valley degeneracy. This
effectively decreases the energy by $k_B T \ln n$, where $n=4$ for one-electron bound states, $n=6$ for two-electron symmetric states 
and $n=10$ for two-electron anti-symmetric states. These numbers are to be changed if the degeneracy lifting due to local lattice effects 
discussed in section \ref{sec:Valley Mixing} or Zeeman splitting exceeds the temperature scale.  
 

An important aspect in graphene is rippling and corrugations \cite{Ripples1,Ripples2,Ripples3}. As argued in ref. 
[\onlinecite{ZeroLLProtection}], corrugations in graphene may be described by fluctuations in perpendicular magnetic field that lead to 
considerable broadening of non-zero LLs. At the same time, the 0-th LL is protected and mainly broadens due to 
temperature \cite{ZeroLLProtection}. 
Our main results correspond to chemical potentials in the gap above or below the 0-th LL, thus the effect of corrugations is expected to
be moderate. At the same time, corrugations may kill any weak binding effects that might occur in the higher QHE plateaux.

To summarize, we have shown that graphene in magnetic field can mediate strong attraction of like charges put near graphene. 
The 
resulting size and configuration of macro-molecules depend on magnetic length and local chemical potential 
and thus can be easily changed by tuning the magnetic field or doping of graphene. The results are expected to survive significant 
temperatures. This opens a perspective to nanoscopic manipulation of ions on graphene by using macroscopic tools and provides further
insight onto the structure of incompressible regions in QHE physics  beyond the semiclassical approximation.  
Apart from that, the results might shed light on the microscopic structure of potential landscape in SiC epitaxial 
Quantum Hall Effect devices.

{\it Acknowledgements:}
This work has been supported by EPSRC through  grants EP/l02669X/1 and EP/H049797/1.  The support of the RSF grant 14-22-00281 is 
acknowledged.   I am indebted to Joseph J. Betouras and Feo 
V. Kusmartsev for many stimulating discussions and support. Useful correspondence with A.Tzalenchuk,  O.Gamayun and G.Berdiyorov is 
gratefully acknowledged. Important comments and suggestions of the Referees are acknowledged.

\bibliography{Graphene}

\begin{thebibliography}{47}%
\makeatletter
\providecommand \@ifxundefined [1]{%
 \@ifx{#1\undefined}
}%
\providecommand \@ifnum [1]{%
 \ifnum #1\expandafter \@firstoftwo
 \else \expandafter \@secondoftwo
 \fi
}%
\providecommand \@ifx [1]{%
 \ifx #1\expandafter \@firstoftwo
 \else \expandafter \@secondoftwo
 \fi
}%
\providecommand \natexlab [1]{#1}%
\providecommand \enquote  [1]{``#1''}%
\providecommand \bibnamefont  [1]{#1}%
\providecommand \bibfnamefont [1]{#1}%
\providecommand \citenamefont [1]{#1}%
\providecommand \href@noop [0]{\@secondoftwo}%
\providecommand \href [0]{\begingroup \@sanitize@url \@href}%
\providecommand \@href[1]{\@@startlink{#1}\@@href}%
\providecommand \@@href[1]{\endgroup#1\@@endlink}%
\providecommand \@sanitize@url [0]{\catcode `\\12\catcode `\$12\catcode
  `\&12\catcode `\#12\catcode `\^12\catcode `\_12\catcode `\%12\relax}%
\providecommand \@@startlink[1]{}%
\providecommand \@@endlink[0]{}%
\providecommand \url  [0]{\begingroup\@sanitize@url \@url }%
\providecommand \@url [1]{\endgroup\@href {#1}{\urlprefix }}%
\providecommand \urlprefix  [0]{URL }%
\providecommand \Eprint [0]{\href }%
\providecommand \doibase [0]{http://dx.doi.org/}%
\providecommand \selectlanguage [0]{\@gobble}%
\providecommand \bibinfo  [0]{\@secondoftwo}%
\providecommand \bibfield  [0]{\@secondoftwo}%
\providecommand \translation [1]{[#1]}%
\providecommand \BibitemOpen [0]{}%
\providecommand \bibitemStop [0]{}%
\providecommand \bibitemNoStop [0]{.\EOS\space}%
\providecommand \EOS [0]{\spacefactor3000\relax}%
\providecommand \BibitemShut  [1]{\csname bibitem#1\endcsname}%
\let\auto@bib@innerbib\@empty
\bibitem [{\citenamefont {{Novoselov}}\ \emph {et~al.}(2005)\citenamefont
  {{Novoselov}}, \citenamefont {{Geim}}, \citenamefont {{Morozov}},
  \citenamefont {{Jiang}}, \citenamefont {{Katsnelson}}, \citenamefont
  {{Grigorieva}}, \citenamefont {{Dubonos}},\ and\ \citenamefont
  {{Firsov}}}]{Geim2005}%
  \BibitemOpen
  \bibfield  {author} {\bibinfo {author} {\bibfnamefont {K.~S.}\ \bibnamefont
  {{Novoselov}}}, \bibinfo {author} {\bibfnamefont {A.~K.}\ \bibnamefont
  {{Geim}}}, \bibinfo {author} {\bibfnamefont {S.~V.}\ \bibnamefont
  {{Morozov}}}, \bibinfo {author} {\bibfnamefont {D.}~\bibnamefont {{Jiang}}},
  \bibinfo {author} {\bibfnamefont {M.~I.}\ \bibnamefont {{Katsnelson}}},
  \bibinfo {author} {\bibfnamefont {I.~V.}\ \bibnamefont {{Grigorieva}}},
  \bibinfo {author} {\bibfnamefont {S.~V.}\ \bibnamefont {{Dubonos}}}, \ and\
  \bibinfo {author} {\bibfnamefont {A.~A.}\ \bibnamefont {{Firsov}}},\ }\href
  {\doibase 10.1038/nature04233} {\bibfield  {journal} {\bibinfo  {journal}
  {\nat}\ }\textbf {\bibinfo {volume} {438}},\ \bibinfo {pages} {197} (\bibinfo
  {year} {2005})},\ \Eprint {http://arxiv.org/abs/cond-mat/0509330}
  {cond-mat/0509330} \BibitemShut {NoStop}%
\bibitem [{\citenamefont {Biggs}\ \emph {et~al.}(2014)\citenamefont {Biggs},
  \citenamefont {Kiamahalleh}, \citenamefont {Mijajlovic},\ and\ \citenamefont
  {Penna}}]{Biggs}%
  \BibitemOpen
  \bibfield  {author} {\bibinfo {author} {\bibfnamefont {M.}~\bibnamefont
  {Biggs}}, \bibinfo {author} {\bibfnamefont {M.}~\bibnamefont {Kiamahalleh}},
  \bibinfo {author} {\bibfnamefont {M.}~\bibnamefont {Mijajlovic}}, \ and\
  \bibinfo {author} {\bibfnamefont {M.}~\bibnamefont {Penna}},\ }\href@noop {}
  {\bibfield  {journal} {\bibinfo  {journal} {AU2014/900273}\ } (\bibinfo
  {year} {2014})}\BibitemShut {NoStop}%
\bibitem [{\citenamefont {Biggs}\ \emph {et~al.}(2015)\citenamefont {Biggs},
  \citenamefont {Penna}, \citenamefont {Kiamahalleh},\ and\ \citenamefont
  {Mijajlovic}}]{Biggs1}%
  \BibitemOpen
  \bibfield  {author} {\bibinfo {author} {\bibfnamefont {M.}~\bibnamefont
  {Biggs}}, \bibinfo {author} {\bibfnamefont {M.}~\bibnamefont {Penna}},
  \bibinfo {author} {\bibfnamefont {M.}~\bibnamefont {Kiamahalleh}}, \ and\
  \bibinfo {author} {\bibfnamefont {M.}~\bibnamefont {Mijajlovic}},\
  }\href@noop {} {\bibfield  {journal} {\bibinfo  {journal}
  {PCT/AU2015/000034}\ } (\bibinfo {year} {2015})}\BibitemShut {NoStop}%
\bibitem [{\citenamefont {Wehling}\ \emph {et~al.}(2009)\citenamefont
  {Wehling}, \citenamefont {Katsnelson},\ and\ \citenamefont
  {Lichtenstein}}]{Wehling2009}%
  \BibitemOpen
  \bibfield  {author} {\bibinfo {author} {\bibfnamefont {T.~O.}\ \bibnamefont
  {Wehling}}, \bibinfo {author} {\bibfnamefont {M.~I.}\ \bibnamefont
  {Katsnelson}}, \ and\ \bibinfo {author} {\bibfnamefont {A.~I.}\ \bibnamefont
  {Lichtenstein}},\ }\href {\doibase 10.1103/PhysRevB.80.085428} {\bibfield
  {journal} {\bibinfo  {journal} {Phys. Rev. B}\ }\textbf {\bibinfo {volume}
  {80}},\ \bibinfo {pages} {085428} (\bibinfo {year} {2009})}\BibitemShut
  {NoStop}%
\bibitem [{\citenamefont {Jin}\ \emph {et~al.}(2010)\citenamefont {Jin},
  \citenamefont {Choi},\ and\ \citenamefont {Jhi}}]{AlkaliGraphene}%
  \BibitemOpen
  \bibfield  {author} {\bibinfo {author} {\bibfnamefont {K.-H.}\ \bibnamefont
  {Jin}}, \bibinfo {author} {\bibfnamefont {S.-M.}\ \bibnamefont {Choi}}, \
  and\ \bibinfo {author} {\bibfnamefont {S.-H.}\ \bibnamefont {Jhi}},\ }\href
  {\doibase 10.1103/PhysRevB.82.033414} {\bibfield  {journal} {\bibinfo
  {journal} {Phys. Rev. B}\ }\textbf {\bibinfo {volume} {82}},\ \bibinfo
  {pages} {033414} (\bibinfo {year} {2010})}\BibitemShut {NoStop}%
\bibitem [{\citenamefont {Khantha}\ \emph {et~al.}(2004)\citenamefont
  {Khantha}, \citenamefont {Cordero}, \citenamefont {Molina}, \citenamefont
  {Alonso},\ and\ \citenamefont {Girifalco}}]{LiGraphene}%
  \BibitemOpen
  \bibfield  {author} {\bibinfo {author} {\bibfnamefont {M.}~\bibnamefont
  {Khantha}}, \bibinfo {author} {\bibfnamefont {N.~A.}\ \bibnamefont
  {Cordero}}, \bibinfo {author} {\bibfnamefont {L.~M.}\ \bibnamefont {Molina}},
  \bibinfo {author} {\bibfnamefont {J.~A.}\ \bibnamefont {Alonso}}, \ and\
  \bibinfo {author} {\bibfnamefont {L.~A.}\ \bibnamefont {Girifalco}},\ }\href
  {\doibase 10.1103/PhysRevB.70.125422} {\bibfield  {journal} {\bibinfo
  {journal} {Phys. Rev. B}\ }\textbf {\bibinfo {volume} {70}},\ \bibinfo
  {pages} {125422} (\bibinfo {year} {2004})}\BibitemShut {NoStop}%
\bibitem [{\citenamefont {{Rudenko}}\ \emph {et~al.}(2010)\citenamefont
  {{Rudenko}}, \citenamefont {{Keil}}, \citenamefont {{Katsnelson}},\ and\
  \citenamefont {{Lichtenstein}}}]{HalogenDFT1}%
  \BibitemOpen
  \bibfield  {author} {\bibinfo {author} {\bibfnamefont {A.~N.}\ \bibnamefont
  {{Rudenko}}}, \bibinfo {author} {\bibfnamefont {F.~J.}\ \bibnamefont
  {{Keil}}}, \bibinfo {author} {\bibfnamefont {M.~I.}\ \bibnamefont
  {{Katsnelson}}}, \ and\ \bibinfo {author} {\bibfnamefont {A.~I.}\
  \bibnamefont {{Lichtenstein}}},\ }\href {\doibase 10.1103/PhysRevB.82.035427}
  {\bibfield  {journal} {\bibinfo  {journal} {\prb}\ }\textbf {\bibinfo
  {volume} {82}},\ \bibinfo {eid} {035427} (\bibinfo {year} {2010})},\ \Eprint
  {http://arxiv.org/abs/1002.2536} {arXiv:1002.2536 [cond-mat.mes-hall]}
  \BibitemShut {NoStop}%
\bibitem [{\citenamefont {{Medeiros}}\ \emph {et~al.}(2010)\citenamefont
  {{Medeiros}}, \citenamefont {{Mascarenhas}}, \citenamefont {{de Brito
  Mota}},\ and\ \citenamefont {{de Castilho}}}]{HalogenDFT}%
  \BibitemOpen
  \bibfield  {author} {\bibinfo {author} {\bibfnamefont {P.~V.~C.}\
  \bibnamefont {{Medeiros}}}, \bibinfo {author} {\bibfnamefont {A.~J.~S.}\
  \bibnamefont {{Mascarenhas}}}, \bibinfo {author} {\bibfnamefont
  {F.}~\bibnamefont {{de Brito Mota}}}, \ and\ \bibinfo {author} {\bibfnamefont
  {C.~M.~C.}\ \bibnamefont {{de Castilho}}},\ }\href {\doibase
  10.1088/0957-4484/21/48/485701} {\bibfield  {journal} {\bibinfo  {journal}
  {Nanotechnology}\ }\textbf {\bibinfo {volume} {21}},\ \bibinfo {eid} {485701}
  (\bibinfo {year} {2010})}\BibitemShut {NoStop}%
\bibitem [{\citenamefont {{Davydov}}\ and\ \citenamefont
  {{Sabirova}}(2011)}]{BondingAdatoms1}%
  \BibitemOpen
  \bibfield  {author} {\bibinfo {author} {\bibfnamefont {S.~Y.}\ \bibnamefont
  {{Davydov}}}\ and\ \bibinfo {author} {\bibfnamefont {G.~I.}\ \bibnamefont
  {{Sabirova}}},\ }\href {\doibase 10.1134/S1063785011060034} {\bibfield
  {journal} {\bibinfo  {journal} {Technical Physics Letters}\ }\textbf
  {\bibinfo {volume} {37}},\ \bibinfo {pages} {515} (\bibinfo {year}
  {2011})}\BibitemShut {NoStop}%
\bibitem [{\citenamefont {{Schedin}}\ \emph {et~al.}(2007)\citenamefont
  {{Schedin}}, \citenamefont {{Geim}}, \citenamefont {{Morozov}}, \citenamefont
  {{Hill}}, \citenamefont {{Blake}}, \citenamefont {{Katsnelson}},\ and\
  \citenamefont {{Novoselov}}}]{GasMovesFreely1}%
  \BibitemOpen
  \bibfield  {author} {\bibinfo {author} {\bibfnamefont {F.}~\bibnamefont
  {{Schedin}}}, \bibinfo {author} {\bibfnamefont {A.~K.}\ \bibnamefont
  {{Geim}}}, \bibinfo {author} {\bibfnamefont {S.~V.}\ \bibnamefont
  {{Morozov}}}, \bibinfo {author} {\bibfnamefont {E.~W.}\ \bibnamefont
  {{Hill}}}, \bibinfo {author} {\bibfnamefont {P.}~\bibnamefont {{Blake}}},
  \bibinfo {author} {\bibfnamefont {M.~I.}\ \bibnamefont {{Katsnelson}}}, \
  and\ \bibinfo {author} {\bibfnamefont {K.~S.}\ \bibnamefont {{Novoselov}}},\
  }\href {\doibase 10.1038/nmat1967} {\bibfield  {journal} {\bibinfo  {journal}
  {Nature Materials}\ }\textbf {\bibinfo {volume} {6}},\ \bibinfo {pages} {652}
  (\bibinfo {year} {2007})}\BibitemShut {NoStop}%
\bibitem [{\citenamefont {{Caragiu}}\ and\ \citenamefont
  {{Finberg}}(2005)}]{ReviewGasMovesFreely2}%
  \BibitemOpen
  \bibfield  {author} {\bibinfo {author} {\bibfnamefont {M.}~\bibnamefont
  {{Caragiu}}}\ and\ \bibinfo {author} {\bibfnamefont {S.}~\bibnamefont
  {{Finberg}}},\ }\href {\doibase 10.1088/0953-8984/17/35/R02} {\bibfield
  {journal} {\bibinfo  {journal} {Journal of Physics Condensed Matter}\
  }\textbf {\bibinfo {volume} {17}},\ \bibinfo {pages} {995} (\bibinfo {year}
  {2005})}\BibitemShut {NoStop}%
\bibitem [{\citenamefont {Shytov}\ \emph {et~al.}(2009)\citenamefont {Shytov},
  \citenamefont {Abanin},\ and\ \citenamefont {Levitov}}]{AbaninClustering08}%
  \BibitemOpen
  \bibfield  {author} {\bibinfo {author} {\bibfnamefont {A.~V.}\ \bibnamefont
  {Shytov}}, \bibinfo {author} {\bibfnamefont {D.~A.}\ \bibnamefont {Abanin}},
  \ and\ \bibinfo {author} {\bibfnamefont {L.~S.}\ \bibnamefont {Levitov}},\
  }\href {\doibase 10.1103/PhysRevLett.103.016806} {\bibfield  {journal}
  {\bibinfo  {journal} {Phys. Rev. Lett.}\ }\textbf {\bibinfo {volume} {103}},\
  \bibinfo {pages} {016806} (\bibinfo {year} {2009})}\BibitemShut {NoStop}%
\bibitem [{\citenamefont {Cheianov}\ \emph {et~al.}(2009)\citenamefont
  {Cheianov}, \citenamefont {Fal’ko}, \citenamefont {Syljuåsen},\ and\
  \citenamefont {Altshuler}}]{CheianovKekule}%
  \BibitemOpen
  \bibfield  {author} {\bibinfo {author} {\bibfnamefont {V.}~\bibnamefont
  {Cheianov}}, \bibinfo {author} {\bibfnamefont {V.}~\bibnamefont {Fal’ko}},
  \bibinfo {author} {\bibfnamefont {O.}~\bibnamefont {Syljuåsen}}, \ and\
  \bibinfo {author} {\bibfnamefont {B.}~\bibnamefont {Altshuler}},\ }\href
  {\doibase http://dx.doi.org/10.1016/j.ssc.2009.07.008} {\bibfield  {journal}
  {\bibinfo  {journal} {Solid State Communications}\ }\textbf {\bibinfo
  {volume} {149}},\ \bibinfo {pages} {1499 } (\bibinfo {year}
  {2009})}\BibitemShut {NoStop}%
\bibitem [{\citenamefont {Goerbig}(2011)}]{ReviewMagnetic}%
  \BibitemOpen
  \bibfield  {author} {\bibinfo {author} {\bibfnamefont {M.~O.}\ \bibnamefont
  {Goerbig}},\ }\href {\doibase 10.1103/RevModPhys.83.1193} {\bibfield
  {journal} {\bibinfo  {journal} {Rev. Mod. Phys.}\ }\textbf {\bibinfo {volume}
  {83}},\ \bibinfo {pages} {1193} (\bibinfo {year} {2011})}\BibitemShut
  {NoStop}%
\bibitem [{\citenamefont {{Slizovskiy}}\ and\ \citenamefont
  {{Betouras}}(2012)}]{Magnetization}%
  \BibitemOpen
  \bibfield  {author} {\bibinfo {author} {\bibfnamefont {S.}~\bibnamefont
  {{Slizovskiy}}}\ and\ \bibinfo {author} {\bibfnamefont {J.~J.}\ \bibnamefont
  {{Betouras}}},\ }\href {\doibase 10.1103/PhysRevB.86.125440} {\bibfield
  {journal} {\bibinfo  {journal} {\prb}\ }\textbf {\bibinfo {volume} {86}},\
  \bibinfo {eid} {125440} (\bibinfo {year} {2012})},\ \Eprint
  {http://arxiv.org/abs/1203.5044} {arXiv:1203.5044 [cond-mat.mes-hall]}
  \BibitemShut {NoStop}%
\bibitem [{\citenamefont {Fogler}\ \emph {et~al.}(2007)\citenamefont {Fogler},
  \citenamefont {Novikov},\ and\ \citenamefont {Shklovskii}}]{screeningFogler}%
  \BibitemOpen
  \bibfield  {author} {\bibinfo {author} {\bibfnamefont {M.~M.}\ \bibnamefont
  {Fogler}}, \bibinfo {author} {\bibfnamefont {D.~S.}\ \bibnamefont {Novikov}},
  \ and\ \bibinfo {author} {\bibfnamefont {B.~I.}\ \bibnamefont {Shklovskii}},\
  }\href {\doibase 10.1103/PhysRevB.76.233402} {\bibfield  {journal} {\bibinfo
  {journal} {Phys. Rev. B}\ }\textbf {\bibinfo {volume} {76}},\ \bibinfo
  {pages} {233402} (\bibinfo {year} {2007})}\BibitemShut {NoStop}%
\bibitem [{\citenamefont {{Terekhov}}\ \emph {et~al.}(2008)\citenamefont
  {{Terekhov}}, \citenamefont {{Milstein}}, \citenamefont {{Kotov}},\ and\
  \citenamefont {{Sushkov}}}]{screening07}%
  \BibitemOpen
  \bibfield  {author} {\bibinfo {author} {\bibfnamefont {I.~S.}\ \bibnamefont
  {{Terekhov}}}, \bibinfo {author} {\bibfnamefont {A.~I.}\ \bibnamefont
  {{Milstein}}}, \bibinfo {author} {\bibfnamefont {V.~N.}\ \bibnamefont
  {{Kotov}}}, \ and\ \bibinfo {author} {\bibfnamefont {O.~P.}\ \bibnamefont
  {{Sushkov}}},\ }\href {\doibase 10.1103/PhysRevLett.100.076803} {\bibfield
  {journal} {\bibinfo  {journal} {Physical Review Letters}\ }\textbf {\bibinfo
  {volume} {100}},\ \bibinfo {eid} {076803} (\bibinfo {year} {2008})},\ \Eprint
  {http://arxiv.org/abs/0708.4263} {arXiv:0708.4263} \BibitemShut {NoStop}%
\bibitem [{\citenamefont {{Pyatkovskiy}}\ and\ \citenamefont
  {{Gusynin}}(2011)}]{Pyatkovskiy}%
  \BibitemOpen
  \bibfield  {author} {\bibinfo {author} {\bibfnamefont {P.~K.}\ \bibnamefont
  {{Pyatkovskiy}}}\ and\ \bibinfo {author} {\bibfnamefont {V.~P.}\ \bibnamefont
  {{Gusynin}}},\ }\href {\doibase 10.1103/PhysRevB.83.075422} {\bibfield
  {journal} {\bibinfo  {journal} {\prb}\ }\textbf {\bibinfo {volume} {83}},\
  \bibinfo {eid} {075422} (\bibinfo {year} {2011})},\ \Eprint
  {http://arxiv.org/abs/1009.5980} {arXiv:1009.5980 [cond-mat.str-el]}
  \BibitemShut {NoStop}%
\bibitem [{\citenamefont {Luican-Mayer}\ \emph {et~al.}(2014)\citenamefont
  {Luican-Mayer}, \citenamefont {Kharitonov}, \citenamefont {Li}, \citenamefont
  {Lu}, \citenamefont {Skachko}, \citenamefont {Gon\ifmmode~\mbox{\c{c}}\else
  \c{c}\fi{}alves}, \citenamefont {Watanabe}, \citenamefont {Taniguchi},\ and\
  \citenamefont {Andrei}}]{Kharitonov}%
  \BibitemOpen
  \bibfield  {author} {\bibinfo {author} {\bibfnamefont {A.}~\bibnamefont
  {Luican-Mayer}}, \bibinfo {author} {\bibfnamefont {M.}~\bibnamefont
  {Kharitonov}}, \bibinfo {author} {\bibfnamefont {G.}~\bibnamefont {Li}},
  \bibinfo {author} {\bibfnamefont {C.-P.}\ \bibnamefont {Lu}}, \bibinfo
  {author} {\bibfnamefont {I.}~\bibnamefont {Skachko}}, \bibinfo {author}
  {\bibfnamefont {A.-M.~B.}\ \bibnamefont {Gon\ifmmode~\mbox{\c{c}}\else
  \c{c}\fi{}alves}}, \bibinfo {author} {\bibfnamefont {K.}~\bibnamefont
  {Watanabe}}, \bibinfo {author} {\bibfnamefont {T.}~\bibnamefont {Taniguchi}},
  \ and\ \bibinfo {author} {\bibfnamefont {E.~Y.}\ \bibnamefont {Andrei}},\
  }\href {\doibase 10.1103/PhysRevLett.112.036804} {\bibfield  {journal}
  {\bibinfo  {journal} {Phys. Rev. Lett.}\ }\textbf {\bibinfo {volume} {112}},\
  \bibinfo {pages} {036804} (\bibinfo {year} {2014})}\BibitemShut {NoStop}%
\bibitem [{\citenamefont {Tetlow}\ \emph {et~al.}(2014)\citenamefont {Tetlow},
  \citenamefont {de~Boer}, \citenamefont {Ford}, \citenamefont {Vvedensky},
  \citenamefont {Coraux},\ and\ \citenamefont {Kantorovich}}]{EpitaxialReview}%
  \BibitemOpen
  \bibfield  {author} {\bibinfo {author} {\bibfnamefont {H.}~\bibnamefont
  {Tetlow}}, \bibinfo {author} {\bibfnamefont {J.~P.}\ \bibnamefont {de~Boer}},
  \bibinfo {author} {\bibfnamefont {I.~J.}\ \bibnamefont {Ford}}, \bibinfo
  {author} {\bibfnamefont {D.~D.}\ \bibnamefont {Vvedensky}}, \bibinfo {author}
  {\bibfnamefont {J.}~\bibnamefont {Coraux}}, \ and\ \bibinfo {author}
  {\bibfnamefont {L.}~\bibnamefont {Kantorovich}},\ }\href {\doibase
  http://dx.doi.org/10.1016/j.physrep.2014.03.003} {\bibfield  {journal}
  {\bibinfo  {journal} {Physics Reports}\ }\textbf {\bibinfo {volume} {542}},\
  \bibinfo {pages} {195} (\bibinfo {year} {2014})}\BibitemShut {NoStop}%
\bibitem [{\citenamefont {Kim}\ \emph {et~al.}(2013)\citenamefont {Kim},
  \citenamefont {Ihm}, \citenamefont {Choi},\ and\ \citenamefont
  {Son}}]{EpitaxialKim}%
  \BibitemOpen
  \bibfield  {author} {\bibinfo {author} {\bibfnamefont {S.}~\bibnamefont
  {Kim}}, \bibinfo {author} {\bibfnamefont {J.}~\bibnamefont {Ihm}}, \bibinfo
  {author} {\bibfnamefont {H.~J.}\ \bibnamefont {Choi}}, \ and\ \bibinfo
  {author} {\bibfnamefont {Y.-W.}\ \bibnamefont {Son}},\ }\href {\doibase
  http://dx.doi.org/10.1016/j.ssc.2013.09.034} {\bibfield  {journal} {\bibinfo
  {journal} {Solid State Communications}\ }\textbf {\bibinfo {volume}
  {175–176}},\ \bibinfo {pages} {83} (\bibinfo {year} {2013})}\BibitemShut
  {NoStop}%
\bibitem [{\citenamefont {Qi}\ \emph {et~al.}(2010)\citenamefont {Qi},
  \citenamefont {Rhim}, \citenamefont {Sun}, \citenamefont {Weinert},\ and\
  \citenamefont {Li}}]{EpitaxialSTM}%
  \BibitemOpen
  \bibfield  {author} {\bibinfo {author} {\bibfnamefont {Y.}~\bibnamefont
  {Qi}}, \bibinfo {author} {\bibfnamefont {S.~H.}\ \bibnamefont {Rhim}},
  \bibinfo {author} {\bibfnamefont {G.~F.}\ \bibnamefont {Sun}}, \bibinfo
  {author} {\bibfnamefont {M.}~\bibnamefont {Weinert}}, \ and\ \bibinfo
  {author} {\bibfnamefont {L.}~\bibnamefont {Li}},\ }\href {\doibase
  10.1103/PhysRevLett.105.085502} {\bibfield  {journal} {\bibinfo  {journal}
  {Phys. Rev. Lett.}\ }\textbf {\bibinfo {volume} {105}},\ \bibinfo {pages}
  {085502} (\bibinfo {year} {2010})}\BibitemShut {NoStop}%
\bibitem [{\citenamefont {Varchon}\ \emph {et~al.}(2007)\citenamefont
  {Varchon}, \citenamefont {Feng}, \citenamefont {Hass}, \citenamefont {Li},
  \citenamefont {Nguyen}, \citenamefont {Naud}, \citenamefont {Mallet},
  \citenamefont {Veuillen}, \citenamefont {Berger}, \citenamefont {Conrad},\
  and\ \citenamefont {Magaud}}]{EpitaxialDeadLayer}%
  \BibitemOpen
  \bibfield  {author} {\bibinfo {author} {\bibfnamefont {F.}~\bibnamefont
  {Varchon}}, \bibinfo {author} {\bibfnamefont {R.}~\bibnamefont {Feng}},
  \bibinfo {author} {\bibfnamefont {J.}~\bibnamefont {Hass}}, \bibinfo {author}
  {\bibfnamefont {X.}~\bibnamefont {Li}}, \bibinfo {author} {\bibfnamefont
  {B.~N.}\ \bibnamefont {Nguyen}}, \bibinfo {author} {\bibfnamefont
  {C.}~\bibnamefont {Naud}}, \bibinfo {author} {\bibfnamefont {P.}~\bibnamefont
  {Mallet}}, \bibinfo {author} {\bibfnamefont {J.-Y.}\ \bibnamefont
  {Veuillen}}, \bibinfo {author} {\bibfnamefont {C.}~\bibnamefont {Berger}},
  \bibinfo {author} {\bibfnamefont {E.~H.}\ \bibnamefont {Conrad}}, \ and\
  \bibinfo {author} {\bibfnamefont {L.}~\bibnamefont {Magaud}},\ }\href
  {\doibase 10.1103/PhysRevLett.99.126805} {\bibfield  {journal} {\bibinfo
  {journal} {Phys. Rev. Lett.}\ }\textbf {\bibinfo {volume} {99}},\ \bibinfo
  {pages} {126805} (\bibinfo {year} {2007})}\BibitemShut {NoStop}%
\bibitem [{\citenamefont {Janssen}\ \emph {et~al.}(2011)\citenamefont
  {Janssen}, \citenamefont {Tzalenchuk}, \citenamefont {Yakimova},
  \citenamefont {Kubatkin}, \citenamefont {Lara-Avila}, \citenamefont
  {Kopylov},\ and\ \citenamefont {Fal'ko}}]{TzalenchukPinning}%
  \BibitemOpen
  \bibfield  {author} {\bibinfo {author} {\bibfnamefont {T.~J. B.~M.}\
  \bibnamefont {Janssen}}, \bibinfo {author} {\bibfnamefont {A.}~\bibnamefont
  {Tzalenchuk}}, \bibinfo {author} {\bibfnamefont {R.}~\bibnamefont
  {Yakimova}}, \bibinfo {author} {\bibfnamefont {S.}~\bibnamefont {Kubatkin}},
  \bibinfo {author} {\bibfnamefont {S.}~\bibnamefont {Lara-Avila}}, \bibinfo
  {author} {\bibfnamefont {S.}~\bibnamefont {Kopylov}}, \ and\ \bibinfo
  {author} {\bibfnamefont {V.~I.}\ \bibnamefont {Fal'ko}},\ }\href {\doibase
  10.1103/PhysRevB.83.233402} {\bibfield  {journal} {\bibinfo  {journal} {Phys.
  Rev. B}\ }\textbf {\bibinfo {volume} {83}},\ \bibinfo {pages} {233402}
  (\bibinfo {year} {2011})}\BibitemShut {NoStop}%
\bibitem [{\citenamefont {{Tzalenchuk}}\ \emph {et~al.}(2010)\citenamefont
  {{Tzalenchuk}}, \citenamefont {{Lara-Avila}}, \citenamefont {{Kalaboukhov}},
  \citenamefont {{Paolillo}}, \citenamefont {{Syv{\"a}j{\"a}rvi}},
  \citenamefont {{Yakimova}}, \citenamefont {{Kazakova}}, \citenamefont
  {{Janssen}}, \citenamefont {{Fal'Ko}},\ and\ \citenamefont
  {{Kubatkin}}}]{TzalenchukNature}%
  \BibitemOpen
  \bibfield  {author} {\bibinfo {author} {\bibfnamefont {A.}~\bibnamefont
  {{Tzalenchuk}}}, \bibinfo {author} {\bibfnamefont {S.}~\bibnamefont
  {{Lara-Avila}}}, \bibinfo {author} {\bibfnamefont {A.}~\bibnamefont
  {{Kalaboukhov}}}, \bibinfo {author} {\bibfnamefont {S.}~\bibnamefont
  {{Paolillo}}}, \bibinfo {author} {\bibfnamefont {M.}~\bibnamefont
  {{Syv{\"a}j{\"a}rvi}}}, \bibinfo {author} {\bibfnamefont {R.}~\bibnamefont
  {{Yakimova}}}, \bibinfo {author} {\bibfnamefont {O.}~\bibnamefont
  {{Kazakova}}}, \bibinfo {author} {\bibfnamefont {T.~J.~B.~M.}\ \bibnamefont
  {{Janssen}}}, \bibinfo {author} {\bibfnamefont {V.}~\bibnamefont {{Fal'Ko}}},
  \ and\ \bibinfo {author} {\bibfnamefont {S.}~\bibnamefont {{Kubatkin}}},\
  }\href {\doibase 10.1038/nnano.2009.474} {\bibfield  {journal} {\bibinfo
  {journal} {Nature Nanotechnology}\ }\textbf {\bibinfo {volume} {5}},\
  \bibinfo {pages} {186} (\bibinfo {year} {2010})},\ \Eprint
  {http://arxiv.org/abs/0909.1220} {arXiv:0909.1220 [cond-mat.mes-hall]}
  \BibitemShut {NoStop}%
\bibitem [{\citenamefont {Alexander-Webber}\ \emph {et~al.}(2013)\citenamefont
  {Alexander-Webber}, \citenamefont {Baker}, \citenamefont {Janssen},
  \citenamefont {Tzalenchuk}, \citenamefont {Lara-Avila}, \citenamefont
  {Kubatkin}, \citenamefont {Yakimova}, \citenamefont {Piot}, \citenamefont
  {Maude},\ and\ \citenamefont {Nicholas}}]{TzalenchukBreakdown}%
  \BibitemOpen
  \bibfield  {author} {\bibinfo {author} {\bibfnamefont {J.~A.}\ \bibnamefont
  {Alexander-Webber}}, \bibinfo {author} {\bibfnamefont {A.~M.~R.}\
  \bibnamefont {Baker}}, \bibinfo {author} {\bibfnamefont {T.~J. B.~M.}\
  \bibnamefont {Janssen}}, \bibinfo {author} {\bibfnamefont {A.}~\bibnamefont
  {Tzalenchuk}}, \bibinfo {author} {\bibfnamefont {S.}~\bibnamefont
  {Lara-Avila}}, \bibinfo {author} {\bibfnamefont {S.}~\bibnamefont
  {Kubatkin}}, \bibinfo {author} {\bibfnamefont {R.}~\bibnamefont {Yakimova}},
  \bibinfo {author} {\bibfnamefont {B.~A.}\ \bibnamefont {Piot}}, \bibinfo
  {author} {\bibfnamefont {D.~K.}\ \bibnamefont {Maude}}, \ and\ \bibinfo
  {author} {\bibfnamefont {R.~J.}\ \bibnamefont {Nicholas}},\ }\href {\doibase
  10.1103/PhysRevLett.111.096601} {\bibfield  {journal} {\bibinfo  {journal}
  {Phys. Rev. Lett.}\ }\textbf {\bibinfo {volume} {111}},\ \bibinfo {pages}
  {096601} (\bibinfo {year} {2013})}\BibitemShut {NoStop}%
\bibitem [{\citenamefont {{Kopylov}}\ \emph {et~al.}(2010)\citenamefont
  {{Kopylov}}, \citenamefont {{Tzalenchuk}}, \citenamefont {{Kubatkin}},\ and\
  \citenamefont {{Fal'ko}}}]{Kopylov}%
  \BibitemOpen
  \bibfield  {author} {\bibinfo {author} {\bibfnamefont {S.}~\bibnamefont
  {{Kopylov}}}, \bibinfo {author} {\bibfnamefont {A.}~\bibnamefont
  {{Tzalenchuk}}}, \bibinfo {author} {\bibfnamefont {S.}~\bibnamefont
  {{Kubatkin}}}, \ and\ \bibinfo {author} {\bibfnamefont {V.~I.}\ \bibnamefont
  {{Fal'ko}}},\ }\href {\doibase 10.1063/1.3487782} {\bibfield  {journal}
  {\bibinfo  {journal} {Applied Physics Letters}\ }\textbf {\bibinfo {volume}
  {97}},\ \bibinfo {eid} {112109} (\bibinfo {year} {2010})},\ \Eprint
  {http://arxiv.org/abs/1007.4340} {arXiv:1007.4340 [cond-mat.mtrl-sci]}
  \BibitemShut {NoStop}%
\bibitem [{\citenamefont {{Slizovskiy}}(2013)}]{MagneticCharging}%
  \BibitemOpen
  \bibfield  {author} {\bibinfo {author} {\bibfnamefont {S.}~\bibnamefont
  {{Slizovskiy}}},\ }\href {\doibase 10.1088/0953-8984/25/49/496007} {\bibfield
   {journal} {\bibinfo  {journal} {Journal of Physics Condensed Matter}\
  }\textbf {\bibinfo {volume} {25}},\ \bibinfo {eid} {496007} (\bibinfo {year}
  {2013})},\ \Eprint {http://arxiv.org/abs/1305.2527} {arXiv:1305.2527
  [cond-mat.mes-hall]} \BibitemShut {NoStop}%
\bibitem [{\citenamefont {Chklovskii}\ \emph {et~al.}(1992)\citenamefont
  {Chklovskii}, \citenamefont {Shklovskii},\ and\ \citenamefont
  {Glazman}}]{ShklovskiiQHE}%
  \BibitemOpen
  \bibfield  {author} {\bibinfo {author} {\bibfnamefont {D.~B.}\ \bibnamefont
  {Chklovskii}}, \bibinfo {author} {\bibfnamefont {B.~I.}\ \bibnamefont
  {Shklovskii}}, \ and\ \bibinfo {author} {\bibfnamefont {L.~I.}\ \bibnamefont
  {Glazman}},\ }\href {\doibase 10.1103/PhysRevB.46.4026} {\bibfield  {journal}
  {\bibinfo  {journal} {Phys. Rev. B}\ }\textbf {\bibinfo {volume} {46}},\
  \bibinfo {pages} {4026} (\bibinfo {year} {1992})}\BibitemShut {NoStop}%
\bibitem [{\citenamefont {Lier}\ and\ \citenamefont
  {Gerhardts}(1994)}]{Gerhardts}%
  \BibitemOpen
  \bibfield  {author} {\bibinfo {author} {\bibfnamefont {K.}~\bibnamefont
  {Lier}}\ and\ \bibinfo {author} {\bibfnamefont {R.~R.}\ \bibnamefont
  {Gerhardts}},\ }\href {\doibase 10.1103/PhysRevB.50.7757} {\bibfield
  {journal} {\bibinfo  {journal} {Phys. Rev. B}\ }\textbf {\bibinfo {volume}
  {50}},\ \bibinfo {pages} {7757} (\bibinfo {year} {1994})}\BibitemShut
  {NoStop}%
\bibitem [{\citenamefont {Koulakov}\ and\ \citenamefont
  {Shklovskii}(1998)}]{ShklovskiiA}%
  \BibitemOpen
  \bibfield  {author} {\bibinfo {author} {\bibfnamefont {A.~A.}\ \bibnamefont
  {Koulakov}}\ and\ \bibinfo {author} {\bibfnamefont {B.~I.}\ \bibnamefont
  {Shklovskii}},\ }\href {\doibase 10.1080/13642819808205015} {\bibfield
  {journal} {\bibinfo  {journal} {Philosophical Magazine Part B}\ }\textbf
  {\bibinfo {volume} {77}},\ \bibinfo {pages} {1235} (\bibinfo {year}
  {1998})},\ \Eprint
  {http://arxiv.org/abs/http://dx.doi.org/10.1080/13642819808205015}
  {http://dx.doi.org/10.1080/13642819808205015} \BibitemShut {NoStop}%
\bibitem [{\citenamefont {Grosberg}\ \emph {et~al.}(2002)\citenamefont
  {Grosberg}, \citenamefont {Nguyen},\ and\ \citenamefont
  {Shklovskii}}]{ShklovskiiBInversion}%
  \BibitemOpen
  \bibfield  {author} {\bibinfo {author} {\bibfnamefont {A.~Y.}\ \bibnamefont
  {Grosberg}}, \bibinfo {author} {\bibfnamefont {T.~T.}\ \bibnamefont
  {Nguyen}}, \ and\ \bibinfo {author} {\bibfnamefont {B.~I.}\ \bibnamefont
  {Shklovskii}},\ }\href {\doibase 10.1103/RevModPhys.74.329} {\bibfield
  {journal} {\bibinfo  {journal} {Rev. Mod. Phys.}\ }\textbf {\bibinfo {volume}
  {74}},\ \bibinfo {pages} {329} (\bibinfo {year} {2002})}\BibitemShut
  {NoStop}%
\bibitem [{\citenamefont {{Novoselov}}\ \emph {et~al.}(2007)\citenamefont
  {{Novoselov}}, \citenamefont {{Jiang}}, \citenamefont {{Zhang}},
  \citenamefont {{Morozov}}, \citenamefont {{Stormer}}, \citenamefont
  {{Zeitler}}, \citenamefont {{Maan}}, \citenamefont {{Boebinger}},
  \citenamefont {{Kim}},\ and\ \citenamefont {{Geim}}}]{roomTQHE}%
  \BibitemOpen
  \bibfield  {author} {\bibinfo {author} {\bibfnamefont {K.~S.}\ \bibnamefont
  {{Novoselov}}}, \bibinfo {author} {\bibfnamefont {Z.}~\bibnamefont
  {{Jiang}}}, \bibinfo {author} {\bibfnamefont {Y.}~\bibnamefont {{Zhang}}},
  \bibinfo {author} {\bibfnamefont {S.~V.}\ \bibnamefont {{Morozov}}}, \bibinfo
  {author} {\bibfnamefont {H.~L.}\ \bibnamefont {{Stormer}}}, \bibinfo {author}
  {\bibfnamefont {U.}~\bibnamefont {{Zeitler}}}, \bibinfo {author}
  {\bibfnamefont {J.~C.}\ \bibnamefont {{Maan}}}, \bibinfo {author}
  {\bibfnamefont {G.~S.}\ \bibnamefont {{Boebinger}}}, \bibinfo {author}
  {\bibfnamefont {P.}~\bibnamefont {{Kim}}}, \ and\ \bibinfo {author}
  {\bibfnamefont {A.~K.}\ \bibnamefont {{Geim}}},\ }\href {\doibase
  10.1126/science.1137201} {\bibfield  {journal} {\bibinfo  {journal}
  {Science}\ }\textbf {\bibinfo {volume} {315}},\ \bibinfo {pages} {1379}
  (\bibinfo {year} {2007})}\BibitemShut {NoStop}%
\bibitem [{\citenamefont {{Zhang}}\ \emph {et~al.}(2005)\citenamefont
  {{Zhang}}, \citenamefont {{Tan}}, \citenamefont {{Stormer}},\ and\
  \citenamefont {{Kim}}}]{QHE2005}%
  \BibitemOpen
  \bibfield  {author} {\bibinfo {author} {\bibfnamefont {Y.}~\bibnamefont
  {{Zhang}}}, \bibinfo {author} {\bibfnamefont {Y.-W.}\ \bibnamefont {{Tan}}},
  \bibinfo {author} {\bibfnamefont {H.~L.}\ \bibnamefont {{Stormer}}}, \ and\
  \bibinfo {author} {\bibfnamefont {P.}~\bibnamefont {{Kim}}},\ }\href
  {\doibase 10.1038/nature04235} {\bibfield  {journal} {\bibinfo  {journal}
  {\nat}\ }\textbf {\bibinfo {volume} {438}},\ \bibinfo {pages} {201} (\bibinfo
  {year} {2005})},\ \Eprint {http://arxiv.org/abs/cond-mat/0509355}
  {cond-mat/0509355} \BibitemShut {NoStop}%
\bibitem [{\citenamefont {Gusynin}\ and\ \citenamefont
  {Sharapov}(2005)}]{QHE2005Gusynin}%
  \BibitemOpen
  \bibfield  {author} {\bibinfo {author} {\bibfnamefont {V.~P.}\ \bibnamefont
  {Gusynin}}\ and\ \bibinfo {author} {\bibfnamefont {S.~G.}\ \bibnamefont
  {Sharapov}},\ }\href {\doibase 10.1103/PhysRevLett.95.146801} {\bibfield
  {journal} {\bibinfo  {journal} {Phys. Rev. Lett.}\ }\textbf {\bibinfo
  {volume} {95}},\ \bibinfo {pages} {146801} (\bibinfo {year}
  {2005})}\BibitemShut {NoStop}%
\bibitem [{\citenamefont {Peres}\ \emph {et~al.}(2006)\citenamefont {Peres},
  \citenamefont {Guinea},\ and\ \citenamefont
  {Castro~Neto}}]{GrapheneDisorder2006}%
  \BibitemOpen
  \bibfield  {author} {\bibinfo {author} {\bibfnamefont {N.~M.~R.}\
  \bibnamefont {Peres}}, \bibinfo {author} {\bibfnamefont {F.}~\bibnamefont
  {Guinea}}, \ and\ \bibinfo {author} {\bibfnamefont {A.~H.}\ \bibnamefont
  {Castro~Neto}},\ }\href {\doibase 10.1103/PhysRevB.73.125411} {\bibfield
  {journal} {\bibinfo  {journal} {Phys. Rev. B}\ }\textbf {\bibinfo {volume}
  {73}},\ \bibinfo {pages} {125411} (\bibinfo {year} {2006})}\BibitemShut
  {NoStop}%
\bibitem [{\citenamefont {Gamayun}\ \emph {et~al.}(2011)\citenamefont
  {Gamayun}, \citenamefont {Gorbar},\ and\ \citenamefont {Gusynin}}]{Gamayun}%
  \BibitemOpen
  \bibfield  {author} {\bibinfo {author} {\bibfnamefont {O.~V.}\ \bibnamefont
  {Gamayun}}, \bibinfo {author} {\bibfnamefont {E.~V.}\ \bibnamefont {Gorbar}},
  \ and\ \bibinfo {author} {\bibfnamefont {V.~P.}\ \bibnamefont {Gusynin}},\
  }\href {\doibase 10.1103/PhysRevB.83.235104} {\bibfield  {journal} {\bibinfo
  {journal} {Phys. Rev. B}\ }\textbf {\bibinfo {volume} {83}},\ \bibinfo
  {pages} {235104} (\bibinfo {year} {2011})}\BibitemShut {NoStop}%
\bibitem [{\citenamefont {Zhang}\ \emph {et~al.}(2012)\citenamefont {Zhang},
  \citenamefont {Barlas},\ and\ \citenamefont {Yang}}]{Semiclassical}%
  \BibitemOpen
  \bibfield  {author} {\bibinfo {author} {\bibfnamefont {Y.}~\bibnamefont
  {Zhang}}, \bibinfo {author} {\bibfnamefont {Y.}~\bibnamefont {Barlas}}, \
  and\ \bibinfo {author} {\bibfnamefont {K.}~\bibnamefont {Yang}},\ }\href
  {\doibase 10.1103/PhysRevB.85.165423} {\bibfield  {journal} {\bibinfo
  {journal} {Phys. Rev. B}\ }\textbf {\bibinfo {volume} {85}},\ \bibinfo
  {pages} {165423} (\bibinfo {year} {2012})}\BibitemShut {NoStop}%
\bibitem [{Note1()}]{Note1}%
  \BibitemOpen
  \bibinfo {note} {This way we may consider a hole in the ``dead layer'' of SiC
  epitaxial graphene as a mobile impurity.}\BibitemShut {Stop}%
\bibitem [{\citenamefont {Fano}\ and\ \citenamefont
  {Ortolani}(1988)}]{2DEGEnergy}%
  \BibitemOpen
  \bibfield  {author} {\bibinfo {author} {\bibfnamefont {G.}~\bibnamefont
  {Fano}}\ and\ \bibinfo {author} {\bibfnamefont {F.}~\bibnamefont
  {Ortolani}},\ }\href {\doibase 10.1103/PhysRevB.37.8179} {\bibfield
  {journal} {\bibinfo  {journal} {Phys. Rev. B}\ }\textbf {\bibinfo {volume}
  {37}},\ \bibinfo {pages} {8179} (\bibinfo {year} {1988})}\BibitemShut
  {NoStop}%
\bibitem [{\citenamefont {Skinner}\ and\ \citenamefont
  {Shklovskii}(2013)}]{2DEGSubstrateEnergy}%
  \BibitemOpen
  \bibfield  {author} {\bibinfo {author} {\bibfnamefont {B.}~\bibnamefont
  {Skinner}}\ and\ \bibinfo {author} {\bibfnamefont {B.~I.}\ \bibnamefont
  {Shklovskii}},\ }\href {\doibase 10.1103/PhysRevB.87.035409} {\bibfield
  {journal} {\bibinfo  {journal} {Phys. Rev. B}\ }\textbf {\bibinfo {volume}
  {87}},\ \bibinfo {pages} {035409} (\bibinfo {year} {2013})}\BibitemShut
  {NoStop}%
\bibitem [{Note2()}]{Note2}%
  \BibitemOpen
  \bibinfo {note} {This will not be the case for the ordinary 2D electron gas
  since the quasiparticle mass will essentially enter the game.}\BibitemShut
  {Stop}%
\bibitem [{\citenamefont {{Castro Neto}}\ \emph {et~al.}(2009)\citenamefont
  {{Castro Neto}}, \citenamefont {{Guinea}}, \citenamefont {{Peres}},
  \citenamefont {{Novoselov}},\ and\ \citenamefont {{Geim}}}]{Review}%
  \BibitemOpen
  \bibfield  {author} {\bibinfo {author} {\bibfnamefont {A.~H.}\ \bibnamefont
  {{Castro Neto}}}, \bibinfo {author} {\bibfnamefont {F.}~\bibnamefont
  {{Guinea}}}, \bibinfo {author} {\bibfnamefont {N.~M.~R.}\ \bibnamefont
  {{Peres}}}, \bibinfo {author} {\bibfnamefont {K.~S.}\ \bibnamefont
  {{Novoselov}}}, \ and\ \bibinfo {author} {\bibfnamefont {A.~K.}\ \bibnamefont
  {{Geim}}},\ }\href {\doibase 10.1103/RevModPhys.81.109} {\bibfield  {journal}
  {\bibinfo  {journal} {Reviews of Modern Physics}\ }\textbf {\bibinfo {volume}
  {81}},\ \bibinfo {pages} {109} (\bibinfo {year} {2009})},\ \Eprint
  {http://arxiv.org/abs/0709.1163} {arXiv:0709.1163} \BibitemShut {NoStop}%
\bibitem [{\citenamefont {Morozov}\ \emph {et~al.}(2006)\citenamefont
  {Morozov}, \citenamefont {Novoselov}, \citenamefont {Katsnelson},
  \citenamefont {Schedin}, \citenamefont {Ponomarenko}, \citenamefont {Jiang},\
  and\ \citenamefont {Geim}}]{Ripples1}%
  \BibitemOpen
  \bibfield  {author} {\bibinfo {author} {\bibfnamefont {S.~V.}\ \bibnamefont
  {Morozov}}, \bibinfo {author} {\bibfnamefont {K.~S.}\ \bibnamefont
  {Novoselov}}, \bibinfo {author} {\bibfnamefont {M.~I.}\ \bibnamefont
  {Katsnelson}}, \bibinfo {author} {\bibfnamefont {F.}~\bibnamefont {Schedin}},
  \bibinfo {author} {\bibfnamefont {L.~A.}\ \bibnamefont {Ponomarenko}},
  \bibinfo {author} {\bibfnamefont {D.}~\bibnamefont {Jiang}}, \ and\ \bibinfo
  {author} {\bibfnamefont {A.~K.}\ \bibnamefont {Geim}},\ }\href {\doibase
  10.1103/PhysRevLett.97.016801} {\bibfield  {journal} {\bibinfo  {journal}
  {Phys. Rev. Lett.}\ }\textbf {\bibinfo {volume} {97}},\ \bibinfo {pages}
  {016801} (\bibinfo {year} {2006})}\BibitemShut {NoStop}%
\bibitem [{\citenamefont {{Meyer}}\ \emph {et~al.}(2007)\citenamefont
  {{Meyer}}, \citenamefont {{Geim}}, \citenamefont {{Katsnelson}},
  \citenamefont {{Novoselov}}, \citenamefont {{Booth}},\ and\ \citenamefont
  {{Roth}}}]{Ripples2}%
  \BibitemOpen
  \bibfield  {author} {\bibinfo {author} {\bibfnamefont {J.~C.}\ \bibnamefont
  {{Meyer}}}, \bibinfo {author} {\bibfnamefont {A.~K.}\ \bibnamefont {{Geim}}},
  \bibinfo {author} {\bibfnamefont {M.~I.}\ \bibnamefont {{Katsnelson}}},
  \bibinfo {author} {\bibfnamefont {K.~S.}\ \bibnamefont {{Novoselov}}},
  \bibinfo {author} {\bibfnamefont {T.~J.}\ \bibnamefont {{Booth}}}, \ and\
  \bibinfo {author} {\bibfnamefont {S.}~\bibnamefont {{Roth}}},\ }\href
  {\doibase 10.1038/nature05545} {\bibfield  {journal} {\bibinfo  {journal}
  {\nat}\ }\textbf {\bibinfo {volume} {446}},\ \bibinfo {pages} {60} (\bibinfo
  {year} {2007})},\ \Eprint {http://arxiv.org/abs/cond-mat/0701379}
  {cond-mat/0701379} \BibitemShut {NoStop}%
\bibitem [{\citenamefont {{Katsnelson}}\ and\ \citenamefont
  {{Novoselov}}(2007)}]{Ripples3}%
  \BibitemOpen
  \bibfield  {author} {\bibinfo {author} {\bibfnamefont {M.~I.}\ \bibnamefont
  {{Katsnelson}}}\ and\ \bibinfo {author} {\bibfnamefont {K.~S.}\ \bibnamefont
  {{Novoselov}}},\ }\href {\doibase 10.1016/j.ssc.2007.02.043} {\bibfield
  {journal} {\bibinfo  {journal} {Solid State Communications}\ }\textbf
  {\bibinfo {volume} {143}},\ \bibinfo {pages} {3} (\bibinfo {year} {2007})},\
  \Eprint {http://arxiv.org/abs/cond-mat/0703374} {cond-mat/0703374}
  \BibitemShut {NoStop}%
\bibitem [{\citenamefont {Giesbers}\ \emph {et~al.}(2007)\citenamefont
  {Giesbers}, \citenamefont {Zeitler}, \citenamefont {Katsnelson},
  \citenamefont {Ponomarenko}, \citenamefont {Mohiuddin},\ and\ \citenamefont
  {Maan}}]{ZeroLLProtection}%
  \BibitemOpen
  \bibfield  {author} {\bibinfo {author} {\bibfnamefont {A.~J.~M.}\
  \bibnamefont {Giesbers}}, \bibinfo {author} {\bibfnamefont {U.}~\bibnamefont
  {Zeitler}}, \bibinfo {author} {\bibfnamefont {M.~I.}\ \bibnamefont
  {Katsnelson}}, \bibinfo {author} {\bibfnamefont {L.~A.}\ \bibnamefont
  {Ponomarenko}}, \bibinfo {author} {\bibfnamefont {T.~M.}\ \bibnamefont
  {Mohiuddin}}, \ and\ \bibinfo {author} {\bibfnamefont {J.~C.}\ \bibnamefont
  {Maan}},\ }\href {\doibase 10.1103/PhysRevLett.99.206803} {\bibfield
  {journal} {\bibinfo  {journal} {Phys. Rev. Lett.}\ }\textbf {\bibinfo
  {volume} {99}},\ \bibinfo {pages} {206803} (\bibinfo {year}
  {2007})}\BibitemShut {NoStop}%
\end{thebibliography}%

\end{document}